\documentclass[11pt]{article}
\usepackage[T1]{fontenc}
\usepackage{array}
\usepackage{amssymb,amsmath,amsthm,amsfonts,epsfig}
\usepackage{esint}
\usepackage{subfig}
\usepackage{natbib}
\usepackage{fullpage}
\usepackage{hyperref}
\usepackage{graphics}
 \usepackage{color}
 
\begin{document}

\title{Adaptive multiple subtraction with wavelet-based complex unary Wiener filters\footnote{Published in Geophysics, \url{http://dx.doi.org/10.1190/geo2011-0318.1}, Vol. 77, pp. V183--V192, Nov.-Dec. 2012}}

\renewcommand{\thefootnote}{\fnsymbol{footnote}} 

\author{Sergi Ventosa, Sylvain Le Roy, Ir\`ene Huard,  Antonio Pica,\\
H\'erald Rabeson, Patrice Ricarte and Laurent Duval}

\maketitle

\begin{abstract}
Adaptive subtraction is a key element in predictive multiple-suppression methods. It minimizes misalignments and amplitude differences between modeled and actual multiples, and thus reduces multiple contamination in the dataset after subtraction.
Due to the high cross-correlation between their waveform, the main challenge resides in attenuating multiples without distorting primaries. As they overlap on a wide frequency range, we  split this wide-band problem  into a set of more tractable narrow-band filter designs, using a 1D complex wavelet frame.
This decomposition enables a single-pass adaptive subtraction via complex, single-sample (unary)  Wiener filters, consistently estimated on overlapping windows in a complex  wavelet transformed domain. Each unary filter compensates amplitude differences within its frequency support, and can correct
small and large misalignment errors through phase and integer delay corrections. This approach greatly simplifies the matching filter estimation and, despite its simplicity,
narrows the gap between 1D and standard adaptive 2D methods on field data.
\end{abstract}

\section{Introduction}
Multiples correspond to unwanted coherent events related to wave field reflection bounces on given surfaces. 
We refer to the comprehensive survey in \citet{Verschuur_D_2006_book_sei_mrtppf} for a 
detailed description. 
Their attenuation \citep{Verschuur_D_1992_j-geophysics_ada_srme}
represents one of the greatest challenges in past and present seismic processing, since the January 1948 issue of Geophysics. 
Their processing mainly
relies on two types of approaches: (1) primary/multiple discrimination based on different velocities and reflections 
and (2) periodicity and predictability.

The first approach maps data to a transformed domain with minimum overlap between primaries and multiples.
Transformed domain filters subsequently attenuate multiples or select primaries of interest. The filtered data is finally 
mapped back to data domain, using an appropriate inverse transform.
Common tools include stacking combined with Normal Move Out (NMO),
homomorphic filtering \citep{Buttkus_B_1975_j-geophys-prospect_hom_ftp}, $f-k$ 
\citep{Wu_M_2011_j-tle_cas_sfkd2dosd}, and 
$\tau-p$ representations \citep{Nuzzo_L_2004_j-geophysics_imp_gprcnatpwt}, or alternative breeds ---  parabolic, hyperbolic  ---    of the Radon transform  \citep{Hampson_D_1986_j-can-j-explor-geophys_inverse_vsme,Trad_D_2003_j-geophysics_lat_vsrt,Nowak_E_2006_j-geophysics_amp_prbmrf}, with  pros \& cons discussion in \citet{Kabir_N_2007_j-cseg-recorder_par_rtmrhaavoa}. 

The second approach includes prediction filters and variations thereof \citep{Taner_M_1980_j-geophys-prospect_lon_psfms,Abma_R_2005_j-tle_com_asmma,Spitz_S_2009_p-eage-marine-w_sim_sswfmpefas}, also termed identification, updating, shaping, sequential or matching filters in electrical engineering literature \citep{Ristow_D_1979_j-geophys-prospect_tim_vpfmu} or predictive deconvolution \citep{Taner_M_1995_j-geophys-prospect_lon_pmspdxtd}. Recently, modeling based techniques, such as surface-related multiple estimation (SRME), have demonstrated excellent performance, allowing data-driven multiple removal \citep{Verschuur_D_1992_j-geophysics_ada_srme,Weglein_A_1997_j-geophysics_inv_ssmamsrd,Lin_D_2004_p-seg_3d_srmeagm} or model-driven, wave-equation based, multiple removal \citep{Pica_A_2005_j-tle_3d_srmm,Weisser_T_2006_j-fb_wav_emmai3dsrme}. These methods, based on multiple models, consist in  predicting then subtracting multiple events from original seismic data. 
 
To compensate for mismatches, an adaptive subtraction is usually performed \citep{Verschuur_D_1997_j-geophysics_est_msiip2pae,Dragoset_B_2010_j-geophysics_per_3dsrme};
however, more direct subtraction, for instance when formulated as an inverse problem, is possible \citep[see, e.g.,][]{Amundsen_L_2001_j-geophysics_multidimensional_sdfsmemmobsd, VanGroenestijn_2009_j-geophysics_estimation_pnorsimda}. The adapted model is commonly obtained through a linear Wiener-type filter \citep{Robinson_E_1967_j-geophys-prospect_pri_dwf} that minimizes the energy of the difference between data and model. Although primaries and  multiples, generated from the same source, are not fully uncorrelated, or orthogonal, least square error (LSE) solutions are widespread, mainly due to their computational simplicity. To reduce the distortion of primaries 
due to their cross-correlation with multiples, these matching filters are usually applied in two steps in practical cases. Long filters compensate global amplitude, waveform and time-shift differences, then shorter filters correct time variant discrepancies.

The observation on primaries and multiples cross-correlation leads to more hybrid methods, mixing both transform and prediction approaches. 
Improvements may reside in different strategies. Either the use of more robust objective matching criteria than LSE based ones, such as  those based on the $L_1$ norm \citep{Guitton_A_2004_j-geophys-prospect_ada_smul1n}, or independence-based subtraction \citep{Kaplan_S_2008_j-geophysics_ada_sfsmica} are efficient alternatives.
Simultaneous estimation of sets of filters based on  a broader quantity of information is another option, e.g., in multichannel adaptation \citep{Wang_Y_2003_j-geophysics_mu_suemmf}.
A suitable data representation may further  serve adaptive subtraction (\citet{Taner_M_1980_j-geophys-prospect_lon_psfms} with radial suppression, \citet{Berkhout_A_2006_j-geophysics_foc_ticsrnr} with focal transform). A recent trend focuses on multiscale or wavelet-like approaches \citep{Jacques_L_2011_j-sp_pan_mgrisdfs}, which may better promote sparsity in exploiting slight differences in primaries and multiple  spectra. \citet{Pokrovskaia_T_2004_p-eage_att_rmcnwtd,Ahmed_I_2007_p-seg_2d_wtdase3dsrme} use standard discrete wavelet transforms. \citet{Herrmann_F_2005_p-eage_rob_cdpmssc,Donno_D_2010_j-geophysics_cur_bmp,Neelamani_R_2010_j-geophysics_ada_scvct} follow  a twist towards curvelets,  sometimes regarded as a local multiscale Radon transform \citep{DeHoop_M_2009_j-inv-prob_sei_igrtctp}.

In many cases, the choice of the type and length of the matching filter, whatever the domain, needs careful parameter selection and testing, and requires computational resources. With our method, we aim at overcoming this issue by decomposing a complicated wide-band problem into a set of more tractable narrow-band problems, through a wavelet transform.
In the proposed approach, model-based multiple removal is solely based on 1D traces, and hence does not compete directly with higher dimensional \citep{Dragoset_B_2010_j-geophysics_per_3dsrme}, regularized \citep{Guitton_A_2004_j-geophys-prospect_pat_bamra3dgmds,Fomel_S_2009_j-geophysics_ada_msrnr} or robust approaches \citep{Herrmann_F_2005_p-eage_rob_cdpmssc}. 
More precisely, comparing with existing literature, we retain the following three concepts: 1) non-stationary Wiener 
matching filters \citep{Lines_L_1996_j-can-j-explor-geophys_sup_spmdmbi,Fomel_S_2009_j-geophysics_ada_msrnr}, 2)  complex trace processing \citep{Monk_D_1993_j-geophys-prospect_wav_emscge,Wang_Y_2003_j-geophysics_mu_suemmf,Huo_S_2009_j-geophysics_imp_assma}, 3) continuous wavelet frames \citep{Sinha_S_2005_j-geophysics_spe_dsdcwt}. 

The adopted complex Morlet wavelet frame emulates limited-scale complex derivatives \citep{Monk_D_1993_j-geophys-prospect_wav_emscge}. The adaptation in the wavelet domain is performed at each scale with a  complex "unary" filter, e.g., a one-coefficient matching operator \citep{Pesquet_J_1997_icipa_new_weid}, accounting for localized phase and amplitude variations between data and model. 
The flexible redundancy and the "complex trace" effect of this wavelet frame allow a better management of time variability in model misalignment errors,
at the price of a less intuitive interpretation of the designed  subtraction operator, due to the complex interplay between unary filters and wavelet decomposition.
The complex curvelet method, proposed by \citet{Neelamani_R_2010_j-geophysics_ada_scvct}, can be viewed as performing unary Wiener filtering in a complex-valued curvelet domain. 
With the proposed algorithm, we aim at a better understanding of the intricate relationship between a sparsifying transform and its associated matched filtering. Its flexibility both resides in choices for a quasi-analytic wavelet and its  scale discretization, to better adapt variations in data character and sampling. 
The algorithm alleviates some constraints of the curvelets (either real or complex) such as a relatively high memory footprint and its fixed dyadic frequency decomposition.  However, as strictly 1D, it does not capitalize the information given by neighboring traces. Despite this,
the closed-form of the matching filter may be beneficial in keeping low complexity, computational speed and possible parallelization. Being nevertheless rather competitive to 2D techniques, it may fit sparsely or non-uniformly sampled datasets.

\section{Methodology}

A classical 1D trace observation model is:
\begin{equation}
d[n]=p[n]+m[n]+w[n]\,,\label{eq:Signals_definition}
\end{equation}
 where $d[n]$, $p[n]$, $m[n]$  and $w[n]$ denote the recorded data, primary events, multiples and background noise, respectively, at discrete time index $n$.

\subsection{Complex wavelet transform decomposition}

We perform  a time-scale decomposition of
each data  $d[n]$ and multiple model $x[n]$ traces  with discrete wavelet frame
approximations to continuous wavelet transform (CWT).
We refer to \citet{Jorgensen_P_2009_incoll_com_dcwt} and references therein for
comparison between discrete and continuous wavelet transforms.
We choose the complex Morlet wavelet since  
it yields a simple interpretation of amplitude and phase delay in
the transformed domain.

The mother Morlet wavelet is   approximately analytic and
writes:\begin{equation}
\psi(t)=\pi^{-1/4}e^{-i\omega_{0}t}e^{-t^{2}/2}\,,\label{eq:Morlet_Wavelet}
\end{equation}
where 
$\omega_{0}$ is the central frequency of the modulated Gaussian, $t$ denoting
the continuous time variable.
A standard choice of $\pi\sqrt{2/\ln(2)}$ for the  central
frequency halves side-lobe amplitude with respect to the main lobes.

The associated discrete family of functions is defined
as a sampling of the mother wavelet:
\begin{equation}
\psi_{r,j}^{v}[n]=\frac{1}{\sqrt{2^{j+v/V}}}\psi\!\left(\frac{nT-r2^{j}b_{0}}{2^{j+v/V}}\right)\,,
\label{eq:Wavelet_Family}
\end{equation}
with $T$ the sampling rate and $V$ the number of voices per octave. Indices  $r,j\in\mathbb{Z}$ and $v\in[0,\ldots,V-1]$ denote, respectively, discretized time, 
octave, and voice. 
Finally, $b_{0}$ stands for the sampling period at scale zero, $j=v=0$.
Its redundancy, of approximately $2V/b_0$, 
may be adapted to the expected computational efficiency.

The time-scale representation of trace $d[n]$, for instance, is given by
the inner product: \begin{equation}
\mathbf{d} = d_{r,j}^{v}=\left\langle d[n],\psi_{r,j}^{v}[n]\right\rangle
=\sum_{n}d[n]\overline{\psi_{r,j}^{v}[n]}\,,\label{eq:DWT}\end{equation}
and written in bold, with indices dropped for more generality, accounting for
the whole trace, or a windowed portion of it.
Windowing in time-scale domain improves the localization in  complex unary filters estimation. Expression
$\overline{\psi_{r,j}^{v}[n]}$ denotes the complex conjugate of $\psi_{r,j}^{v}[n]$.

Suppose $\hat{\mathbf{d}}$ results from some time-scale correction or processing of $\mathbf{d}$, here 
model matching and subtraction. Then the resulting filtered trace is synthesized
back to the time domain 
as a sum of the dual frame components, $\widetilde{\psi}_{r,j } ^{v}[n] $,
\begin{equation}\hat{d}[n]=\sum_{r}\sum_{j,v}\hat{d}_{r,j}^{v}\widetilde{
\psi}_{r,j } ^{v}[n] \,,\label{eq:IFWT}\end{equation}
 weighted by the corrected signal $\hat{d}_{r,j}^{v}$, for instance here with multiples subtracted.
Illustrations of corresponding time-scale raw and processed data are provided in \cite{Ventosa_S_2011_p-eage_complex_wamsuf}.
With sufficient redundancy, $\widetilde{\psi}[n]$ can be approximated by $\psi[n]$ up to a constant multiplicative 
factor  \citep[see, e.g.,][]{Vetterli_M_1995_book_wav_sc}.

The discriminative power of the wavelet frame helps reformulate the design of a
long matching filter into combined global and local optimum complex
 unary filters, minimizing the error between
multiple events and their matched model. The straightforward group-delay
estimation allowed by Morlet wavelets, together with 
frame redundancy, helps avoiding reconstruction artifacts observed
with non-redundant discrete wavelet processing
\citep{Gilloire_V_1992_j-ieee-tsp_ada_fsbcsaeaaec}.

\subsection{Unary filter estimation: amplitude and delay}

When the group delay difference between model and multiple sequences is less 
than half a period at all scales, it can be corrected
in the time-scale domain following an LSE
approach. The optimum unary filter at a given wavelet
scale, either in local or global windowed portions of the trace, is defined as
the complex scalar $a$ which,
multiplied by the time-scale decomposed multiple 
model $\mathbf{x}$, minimizes the mean square error $\xi(a)$ 
with the decomposed input dataset $\mathbf{d}$:\begin{equation}
a_{\textrm{opt}}=\mathop{\arg\min}_{a}\xi(a)=\mathop{\arg\min}_{a}\left\Vert \mathbf{d}-a\mathbf{x}\right\Vert ^{2}\label{eq:Min_criterium}\end{equation}
where the phase of
$a_{\textrm{opt}}$ compensates small delays.

This method is efficient, provided the above small group delay condition is satisfied \citep{Neelamani_R_2010_j-geophysics_ada_scvct}.
Otherwise, i.e., when the cross-correlation peak departs from the zero delay 
sample, faulty $a_{\textrm{opt}}$ estimation in equation \ref{eq:Min_criterium} 
requires a redesign of the filter.

Two options can be conceived to add robustness against high fractional
delays, while keeping with the unary filter approach: either to design
filters with a high group delay, or to add a gross group delay
correction first and then to apply the previous unary filter for fine
group delay correction. In contrast with the Fourier transform,
wavelets' fast decay strongly limits the maximum group delay correction 
on the first option. Hence, we focus exclusively on the second one. 
However, high group delay filters enable a relatively simple scale-wise 
unary filter design, that could reinforce matching in noisy scenarios, 
in applications where the specifications of group-delay range are smaller 
than the effective time support of the chosen wavelets.

We thus introduce a delay term into the above unary filter to yield higher 
delay robustness, while keeping an LSE 
criterion. This amounts to finding optimum  $a$ and delay $l$ that minimize: 
\begin{equation}
\xi(a,l)=\sum_{r}\left|d_{j}^{v}[r]-a_{j}^{v}x_{j}^{v}[r-l]\right|^{2}=\left\Vert \mathbf{d}-a\mathbf{x}_{l}\right\Vert ^{2}\,,\label{eq:MMSE_withDelay}\end{equation} 
where $\mathbf{x}_{l}$ denotes a delay of $\mathbf{x}$ by $l$ samples, 
sequences $d_{j}^{v}[r]$ and $x_{j}^{v}[r]$ are complex and 
the $\sum$ symbol denotes a locally weighted sum along odd $L$ consecutive 
samples around index $r$. We opt to drop all subscripts for $a_{\textrm{opt}}$ 
for clarity. It is understood that this algorithm may be applied either locally 
or globally, depending on problem constraints.

For a given delay $l$, optimum $a_{\textrm{opt}}$ yields orthogonality between filtered signal 
$\mathbf{y}=\mathbf{d}-a_{\textrm{opt}}\mathbf{x}_{l}$ and adapted model $a\mathbf{x}_{l}$, 
which writes in inner product form as:\begin{equation}
\left\langle \mathbf{d}-a_{\textrm{opt}}\mathbf{x}_{l},a\mathbf{x}_{l}\right\rangle =0\quad\forall a\neq0\,.\label{eq:Min_Orthogonality}\end{equation}
By developing and applying linearity on the first argument and conjugate 
linearity on the second one, we obtain: \begin{equation}
\overline{a}\left(\left\langle \mathbf{d},\mathbf{x}_{l}\right\rangle -a_{\textrm{opt}}\left\langle \mathbf{x}_{l},\mathbf{x}_{l}\right\rangle \right)=0\quad\forall a\neq0\label{eq:Min_Dev1}\end{equation}
and therefore\begin{equation}
a_{\textrm{opt}}[l]=\frac{\left\langle \mathbf{d},\mathbf{x}_{l}\right\rangle }{\left\Vert \mathbf{x}_{l}\right\Vert ^{2}}=\frac{\sum_{r}d_{j}^{v}[r]\overline{x_{j}^{v}[r-l]}}{\sum_{r}\left|x_{j}^{v}[r-l]\right|^{2}}\,,\label{eq:MMSE_aopt}\end{equation}
i.e., the unary Wiener filter solution for complex signals. 

Note that instead of performing a direct measurement of the fractional delay, or equivalently
of the group delay, we estimate the phase delay at each
scale or frequency component. When the signal-to-noise ratio (S/N) 
is high, all meaningful phase components are sufficiently well
estimated. However, when the S/N decreases, moderate phase
errors made at key scale components may induce huge errors on 
group delay. As a consequence, we have to ensure that the phase 
of $a_{\textrm{opt}}$ varies smoothly in scale for important components.

Several criteria, well adapted to the nature of the seismic signals,
can be chosen in the selection of the optimum delay. 
The optimum  integer delay, in LSE sense, is selected among all sub-optimal $a_{\textrm{opt}}[l]$ as the one minimizing  $\xi(a,l)$:
\begin{equation}
l_{\textrm{opt}}=\mathop{\arg\min}_{l}\xi(a_{\textrm{opt}}[l])\,.\label{eq:IntDelay_MMSE}\end{equation}
Alternatively, due to the importance of waveform
over amplitude, we can define the
optimum delay as the one that maximizes the normalized cross-correlation
between the adapted model and the data, commonly called coherence
\citep{Neidell_N_1971_j-geophysics_sem_ocmmd,Taner_M_1979_j-geophysics_com_sta,Schimmel_M_1997_j-geophys-j-int_noi_rdwcstpws},\begin{equation}
l_{\textrm{opt}}=\mathop{\arg\max}_{l}\: C[l]\label{eq:IntDelay_Coh}\end{equation}
being \begin{equation}
C[l]=\textrm{Re}\left[\frac{\left\langle \mathbf{d},a_{\textrm{opt}}[l]\mathbf{x}_{l}\right\rangle }{\left\Vert \mathbf{d}\right\Vert \left\Vert a_{\textrm{opt}}[l]\mathbf{x}_{l}\right\Vert }\right]\,,\label{eq:IntDelay_nxcorr}\end{equation}
where we focus on the real part of the normalized cross-correlation.

The use of unary filters enables two important simplifications
of the previous equations. Applying conjugate linearity
in the second argument of the inner product and positive scalability
of the norm gives:\begin{equation}
l_{\textrm{opt}}=\mathop{\arg\max}_{l}\:\textrm{Re}\left[\frac{\overline{a_{\textrm{opt}}[l]}}{\left|a_{\textrm{opt}}[l]\right|}\frac{\left\langle \mathbf{d},\mathbf{x}_{l}\right\rangle }{\left\Vert \mathbf{d}\right\Vert \left\Vert \mathbf{x}_{l}\right\Vert }\right]\,.\label{eq:nxcorr_simpl}\end{equation}
Namely, the normalized cross-correlation between the adapted
model and the data only depends on the phase of the optimum unary
filter and the normalized cross-correlation between the original model
and the data.

Slighty better results can be obtained with an independent fine selection of the window functions used in the estimation of the unary filter and the normalized cross-correlation, equations \ref{eq:MMSE_aopt} and \ref{eq:IntDelay_nxcorr}. However,
even more important to us, when these weighted sums 
are chosen equal, the normalized cross-correlation of the corrected sequence
is equivalent to the cosine of the angle between $\mathbf{d}$ and
$\mathbf{x}_{l}$. This can be shown noticing that $\overline{a_{\textrm{opt}}[l]}/\left|a_{\textrm{opt}}[l]\right|$
can be written depending on
$\mathbf{d}$ and $\mathbf{x}_{l}$ from
its definition in
equation \ref{eq:MMSE_aopt},\begin{equation}
\frac{\overline{a_{\textrm{opt}}[l]}}{\left|a_{\textrm{opt}}[l]\right|}=\frac{\overline{\left\langle \mathbf{d},\mathbf{x}_{l}\right\rangle }}{\left|\left\langle \mathbf{d},\mathbf{x}_{l}\right\rangle \right|}\:.\label{eq:phase_aopt}\end{equation}
As a consequence, the normalized cross-correlation is totally independent
of the optimum unary filter in this particular case,\begin{equation}
l_{\textrm{opt}}=\mathop{\arg\max}_{l}\frac{\left|\left\langle \mathbf{d},\mathbf{x}_{l}\right\rangle \right|}{\left\Vert \mathbf{d}\right\Vert \left\Vert \mathbf{x}_{l}\right\Vert }\label{eq:nxcorr_aopt-indep}\end{equation}
and the coherence value measured is always real and positive. 
When no window is used, this criterion reduces to  maximum cross-correlation between data and predicted multiples.

This configuration enables an important reduction on the computational cost
when the normalized cross-correlation estimator is used, as it enables
a direct estimation of the optimum integer delay, 
without having to estimate the optimum unary filter for each candidate.
Concisely, we first gauge an integer delay, equation \ref{eq:nxcorr_aopt-indep}, to rectify large time shifts; we then estimate the optimum unary filter, equation \ref{eq:MMSE_aopt}, to shrink amplitude and remaining delay differences.

\section{Examples}
\begin{figure*}
\includegraphics[width=0.325\textwidth]{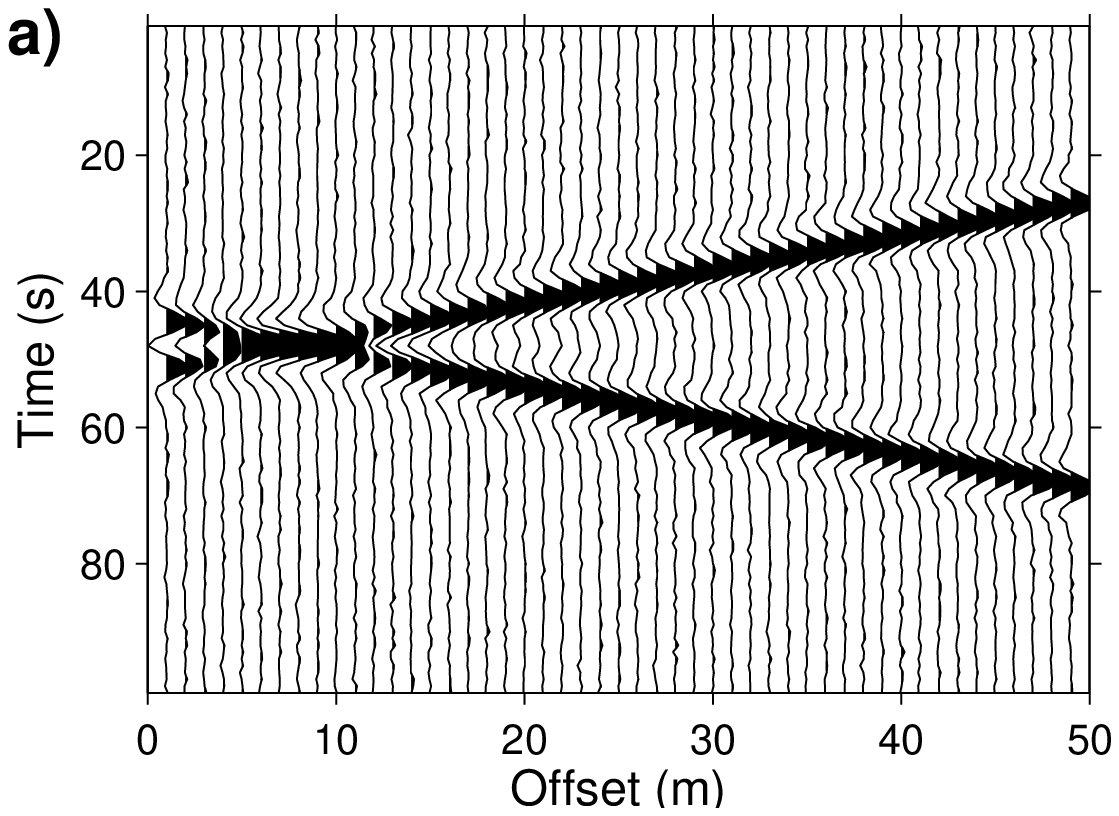} \includegraphics[width=0.325\textwidth]{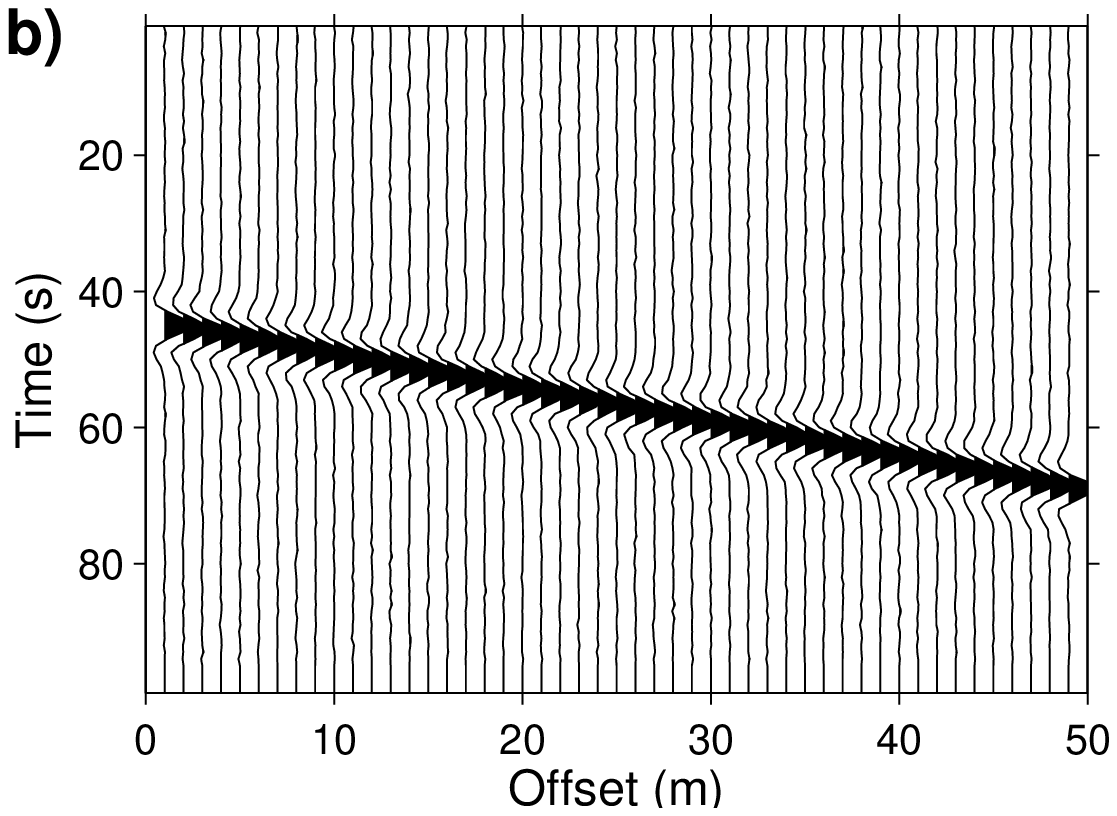}
\includegraphics[width=0.325\textwidth]{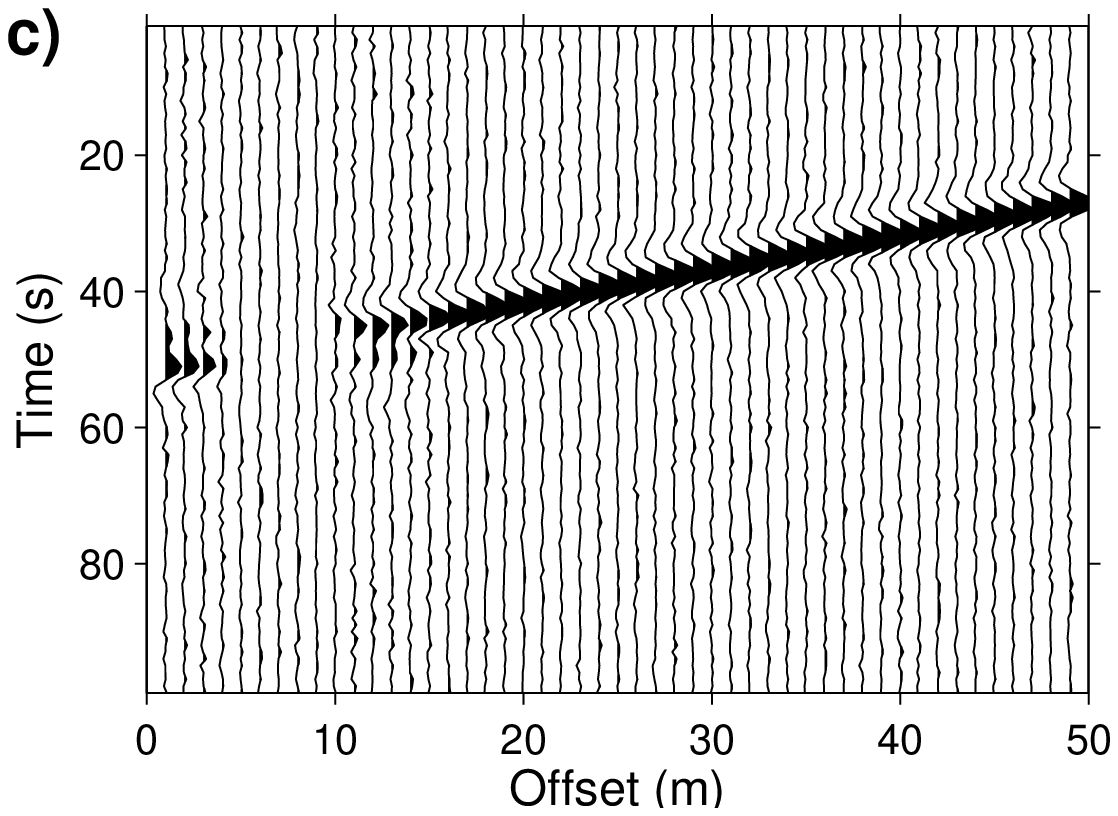} \includegraphics[width=0.325\textwidth]{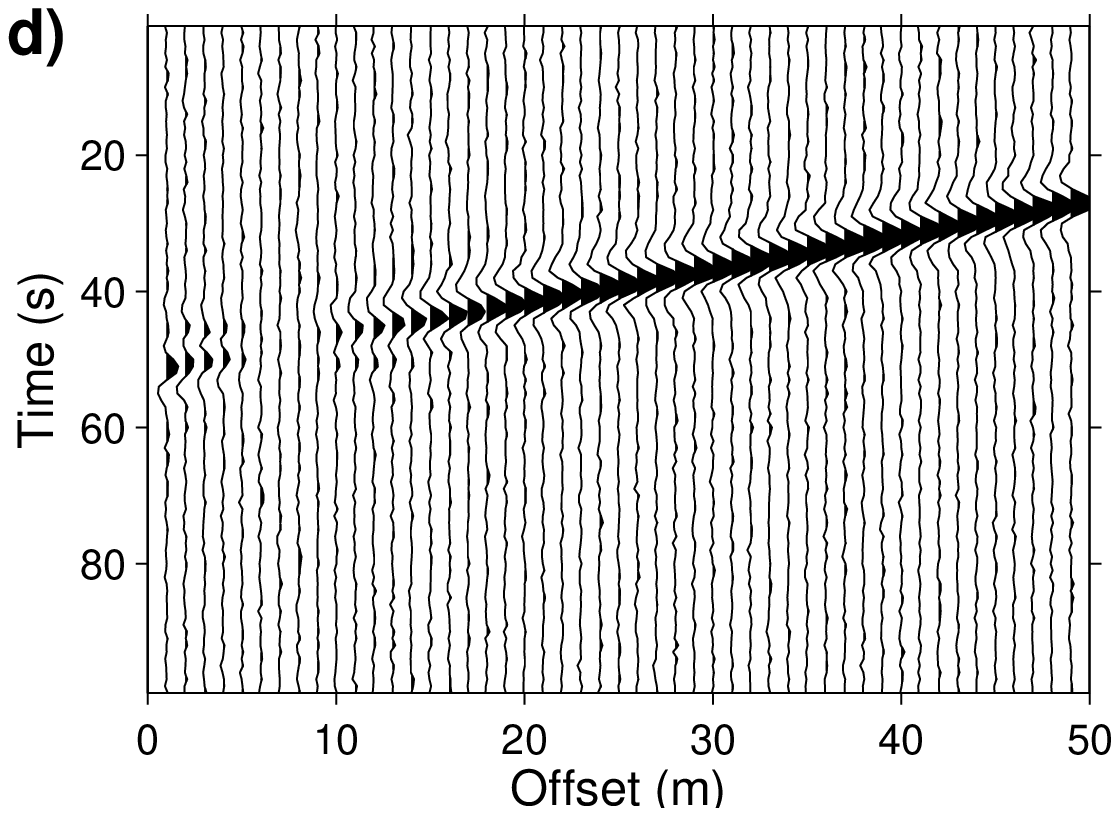}
\includegraphics[width=0.325\textwidth]{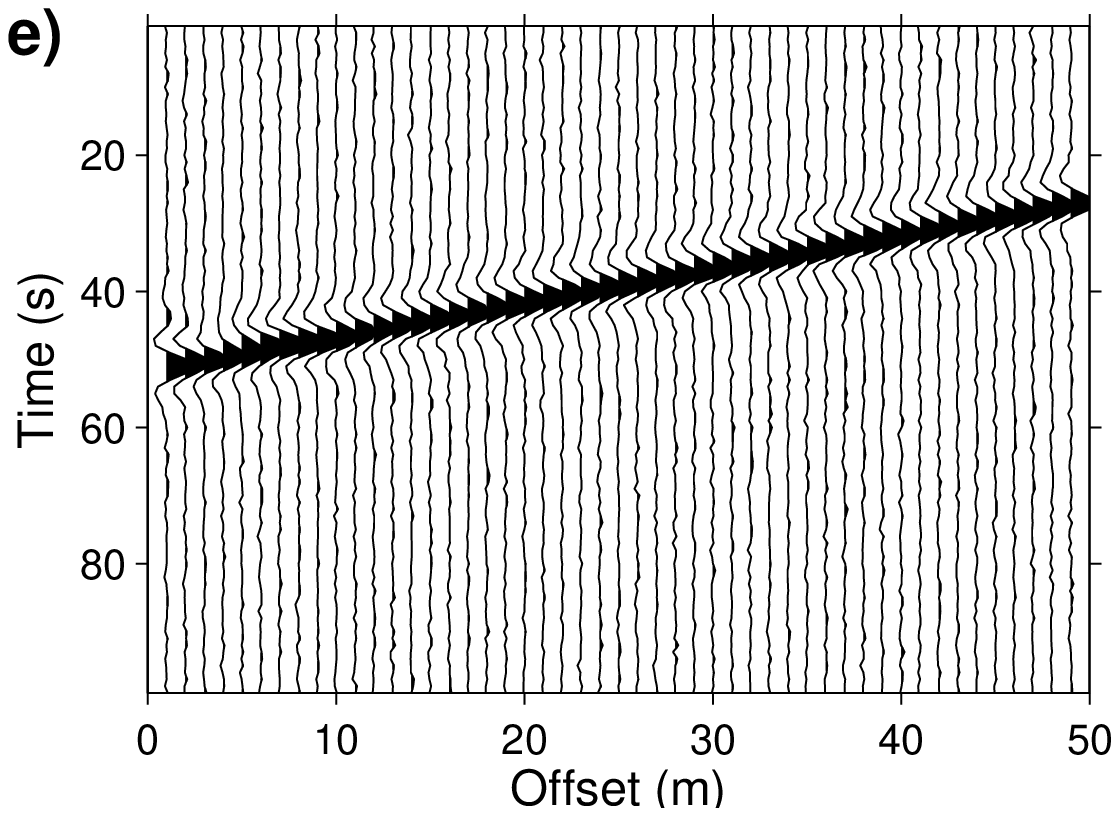} 
\caption{\label{DM_CSP_SYNTH} Subtraction results on a synthetic dataset with two events (a) with multiple model (b), using: (c) standard 1D adaptation, (d) 1D  complex unary filters in time-scale, and (e) standard 2D adaptation.}
\end{figure*}
\begin{figure}
\begin{centering}
\includegraphics[width=0.25\textwidth]{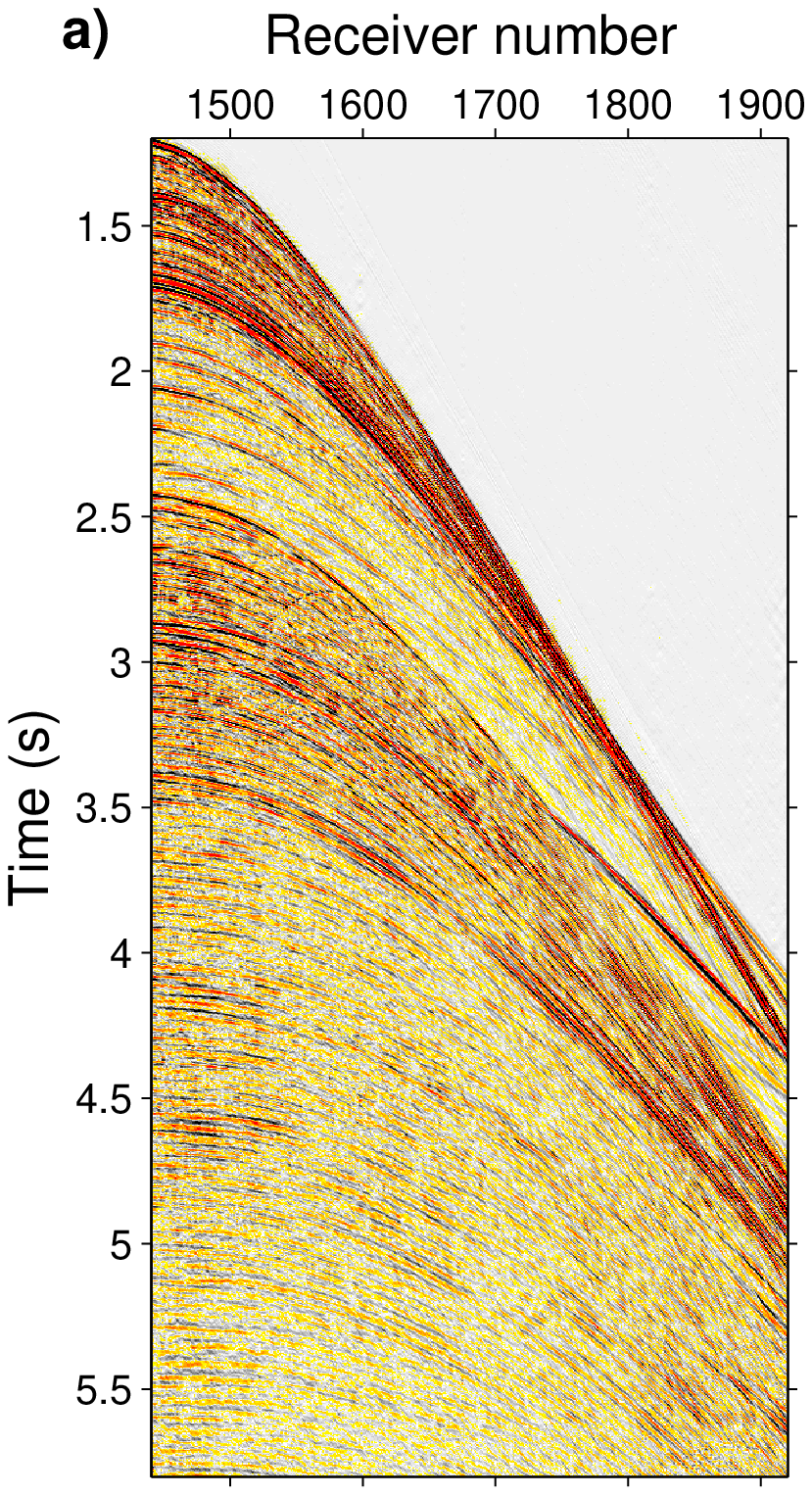}\includegraphics[width=0.231667\textwidth]{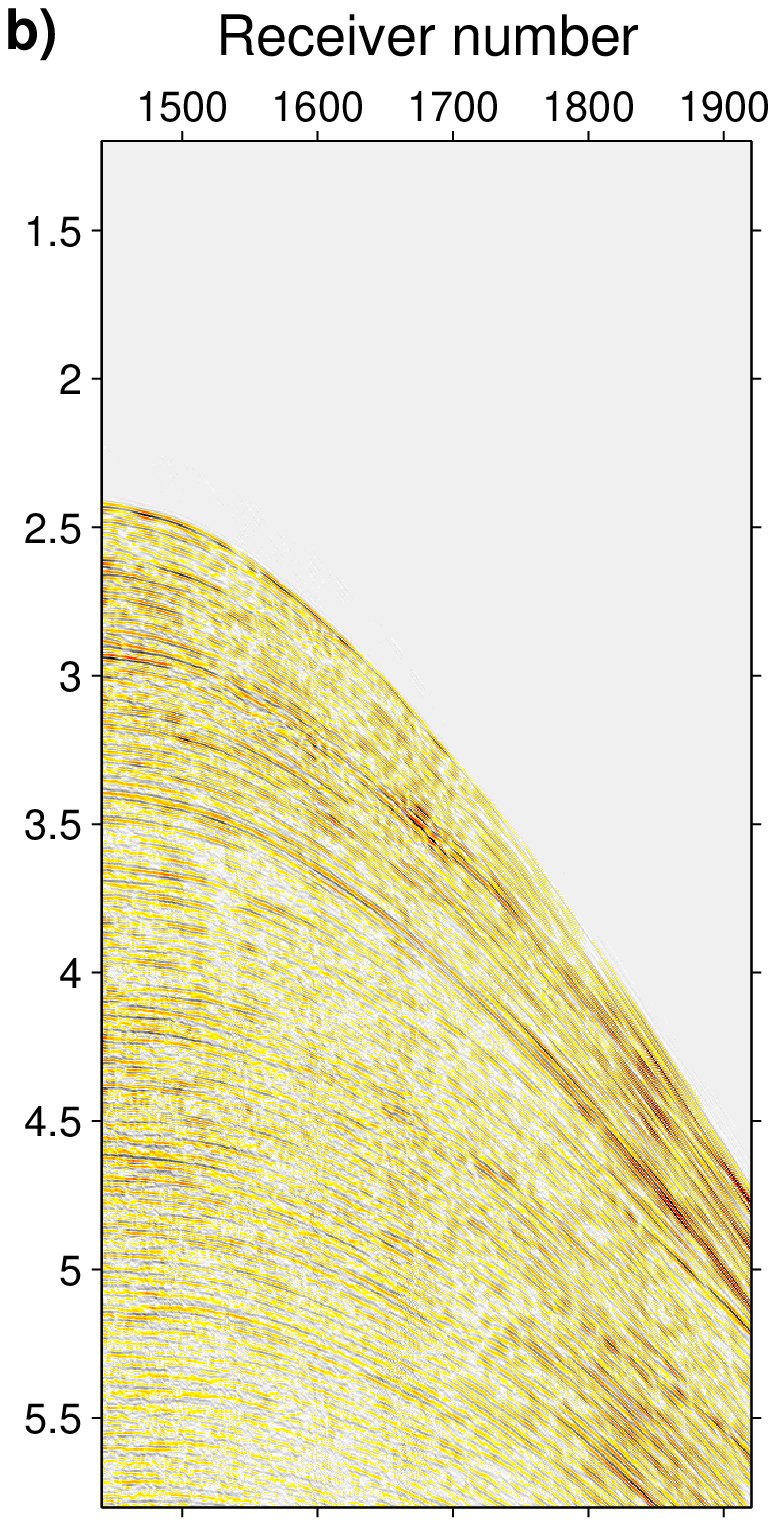}
\par\end{centering}
\caption{\label{fig:DM_CSP}Portion of common shot gather 1551. (a) Recorded data. (b) Multiple model.}
\end{figure}

\begin{figure*}
\includegraphics[width=0.98\textwidth]{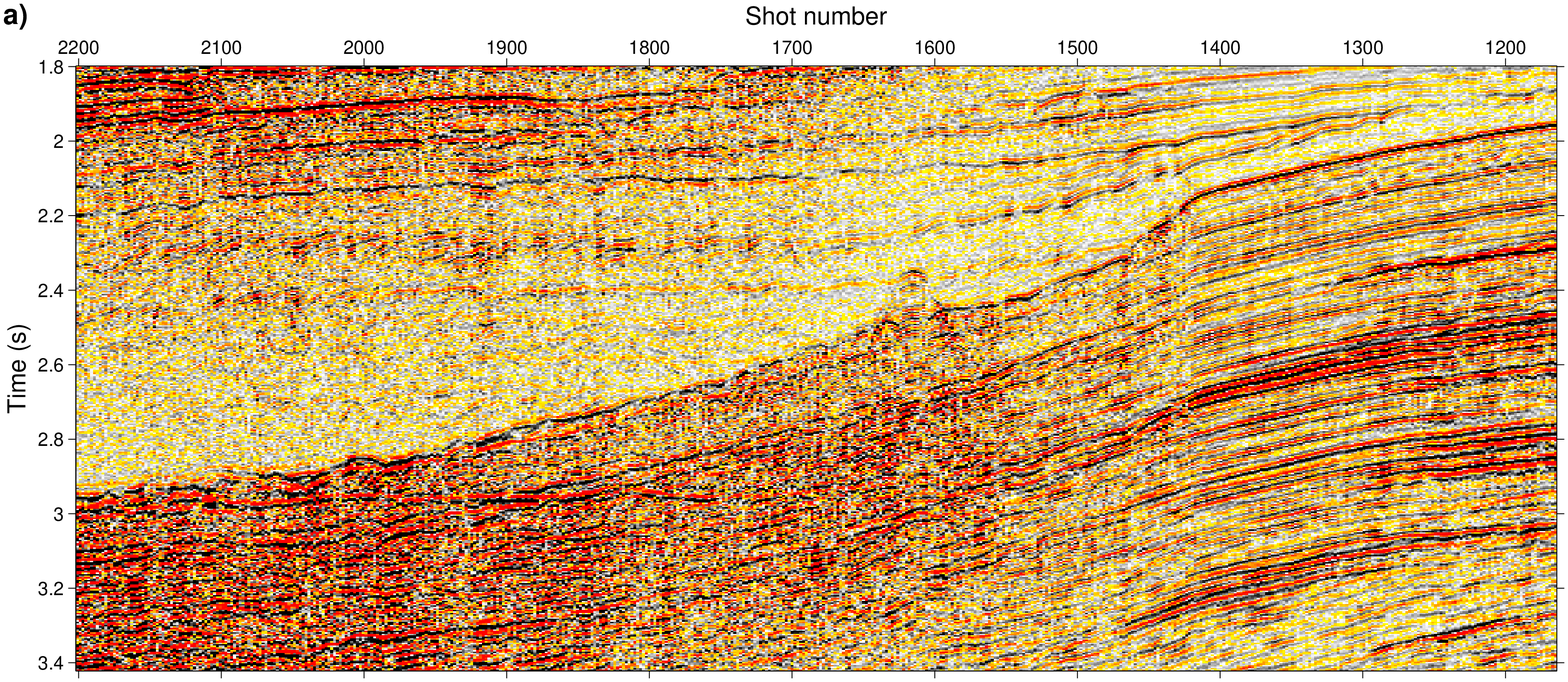}
\includegraphics[width=0.98\textwidth]{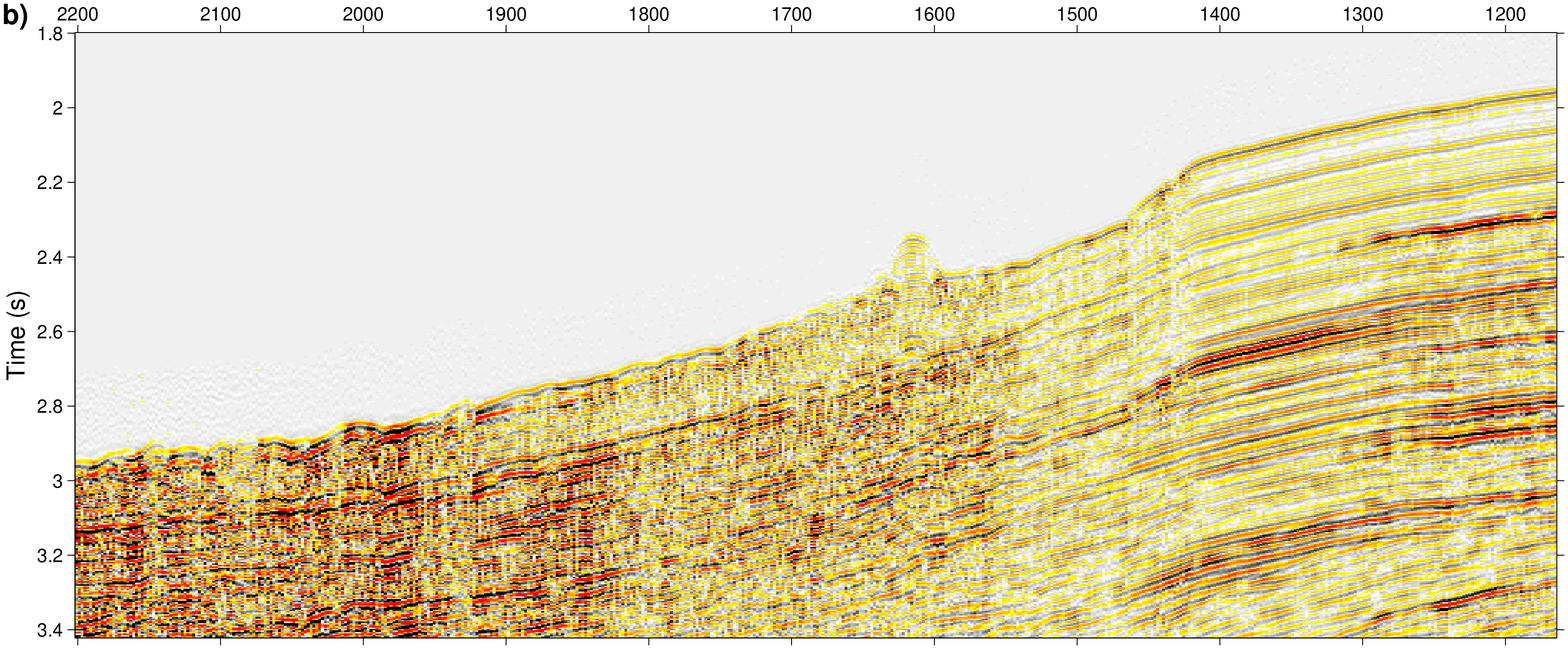}
\caption{\label{fig:DM_CRP1}Portion of first channel gather. (a) Recorded data. (b) Multiple model.}
\end{figure*}

\begin{figure*}
\begin{centering}
\includegraphics[width=0.25\textwidth]{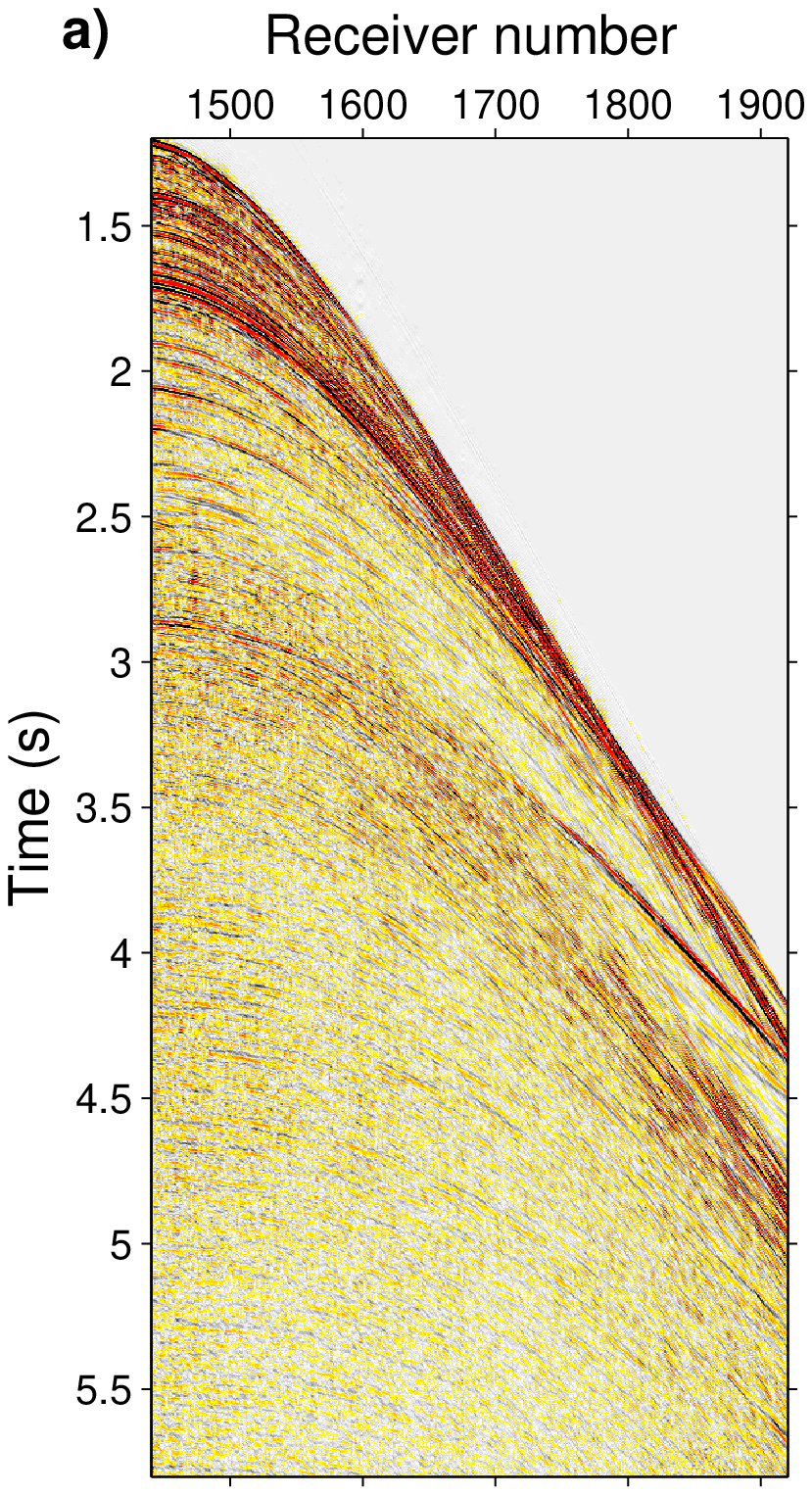}
\includegraphics[width=0.231667\textwidth]{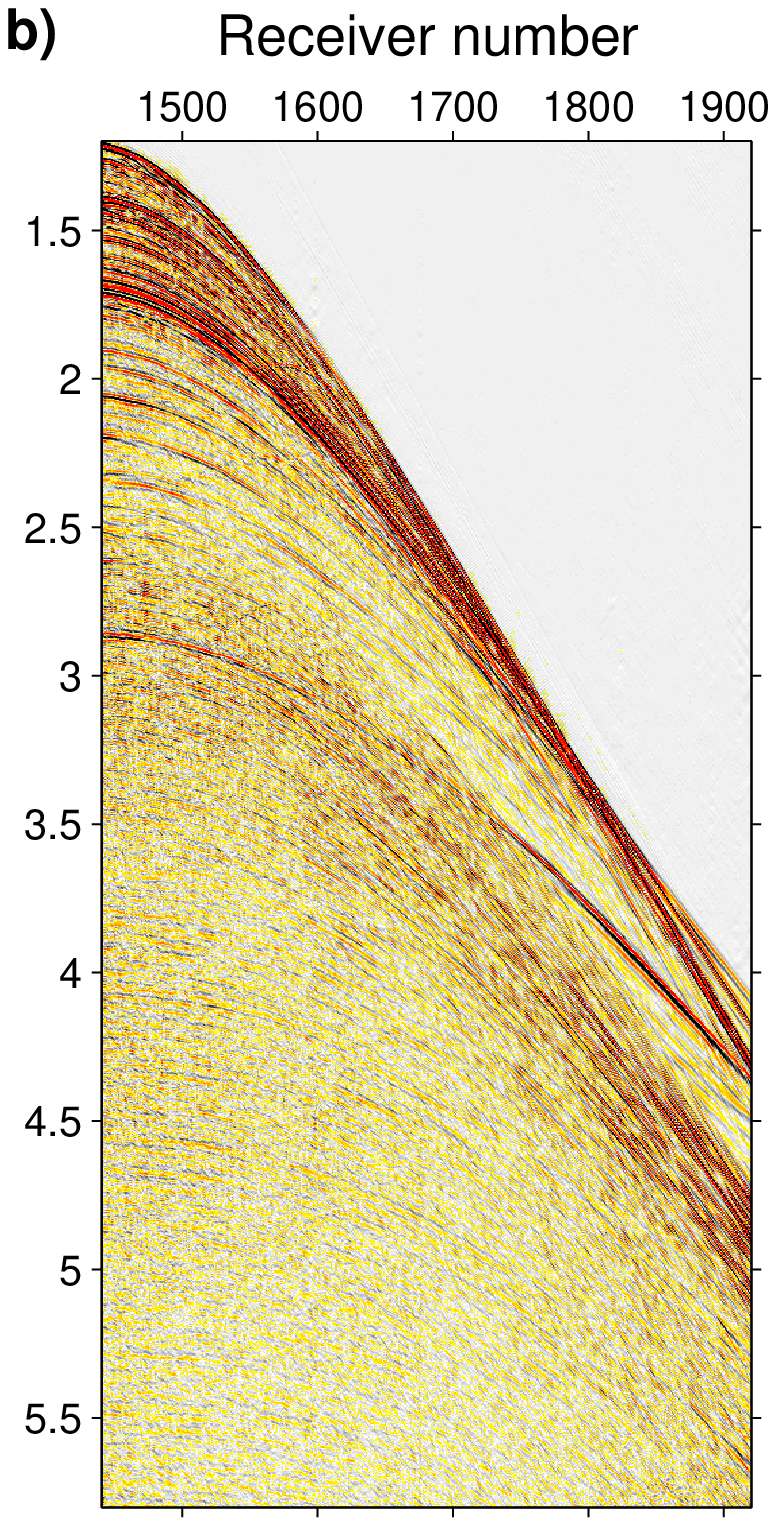}
\includegraphics[width=0.231667\textwidth]{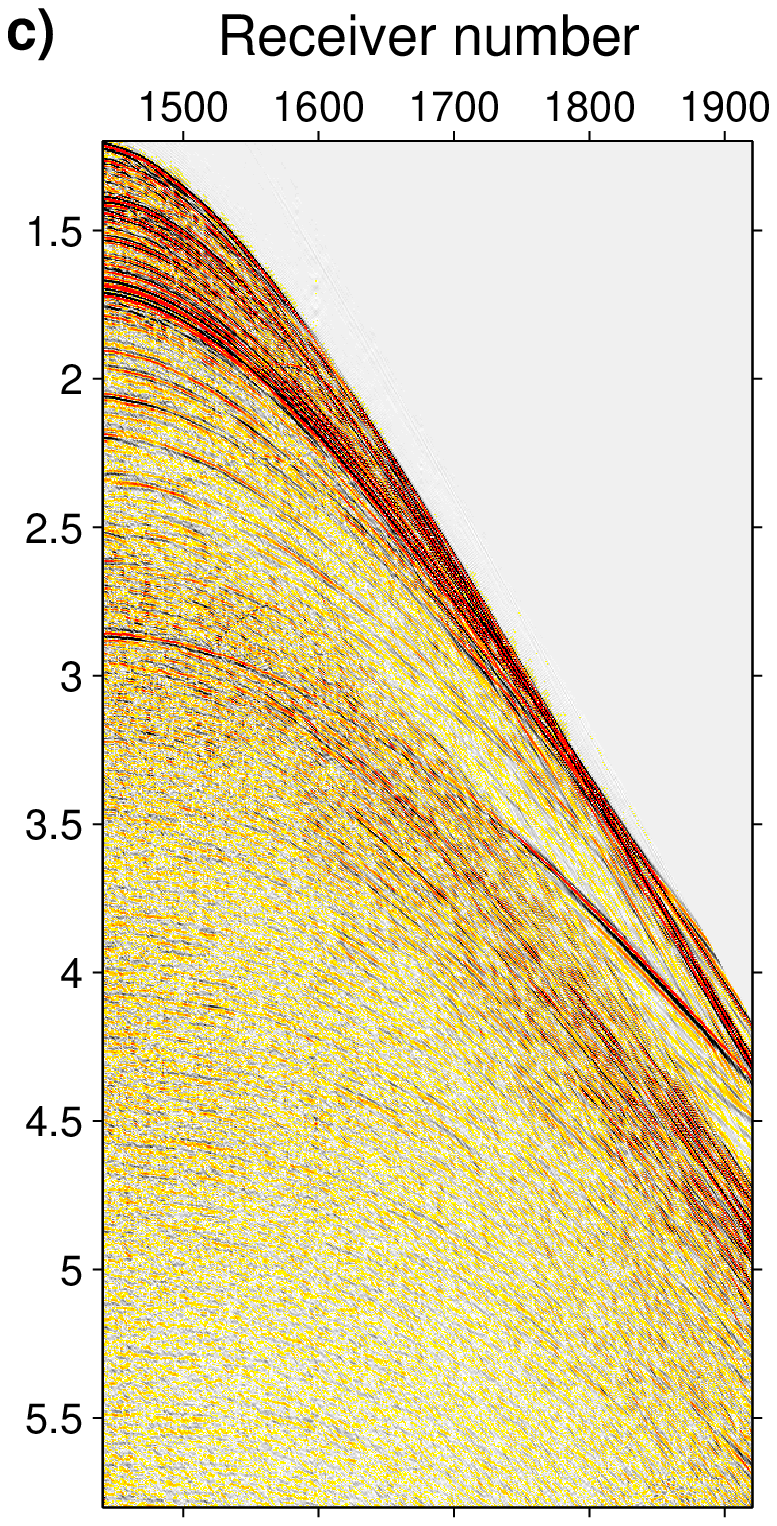}
\includegraphics[width=0.231667\textwidth]{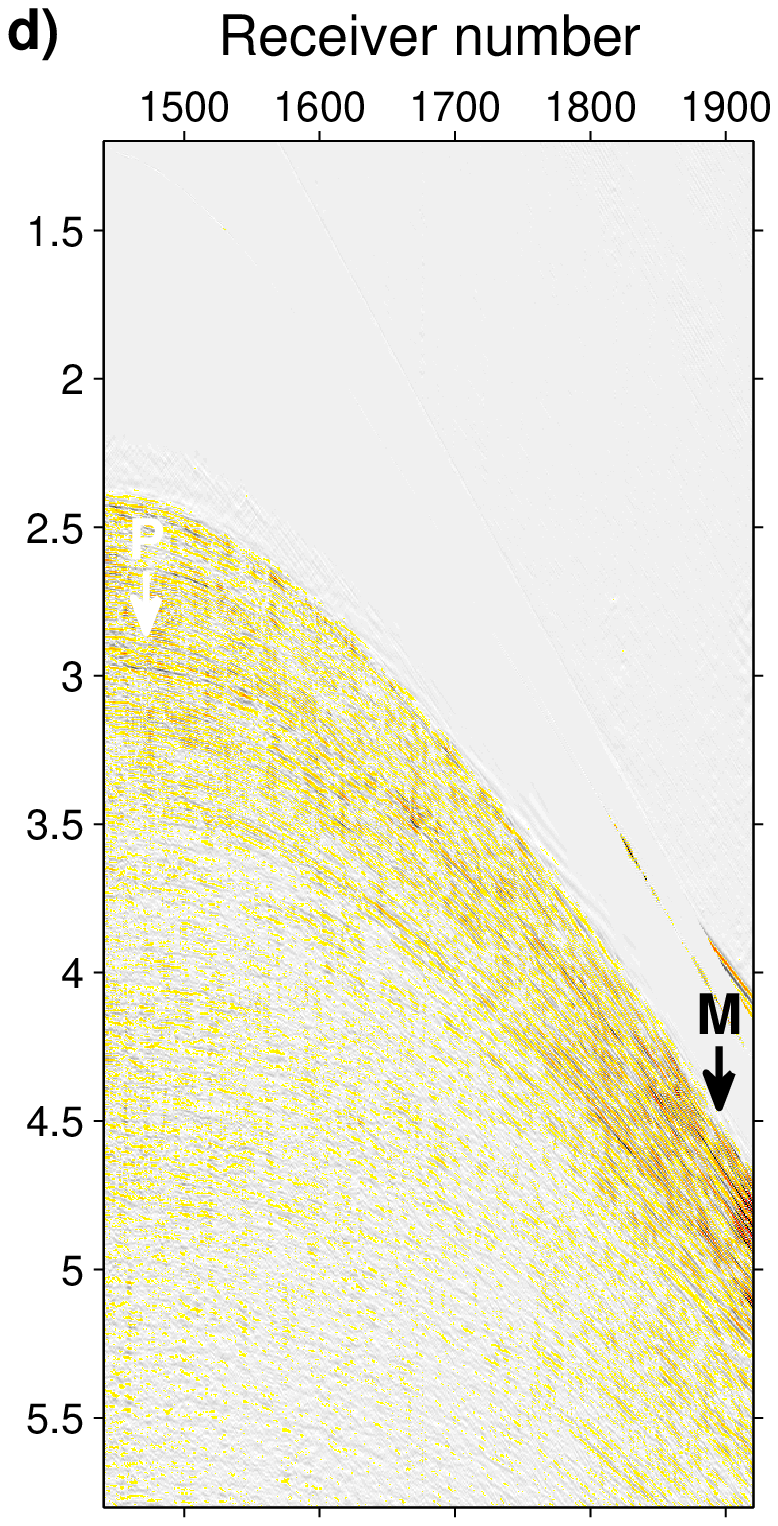}
\par\end{centering}
\caption{\label{fig:DM_FW_FA_CSP}Subtraction results on  common shot gather
1551. Filtered data (a) with the standard 1D adaptive filter, (b) with 1D complex unary filters in the time-scale domain, and (c) with standard 2D adaptive filter in the time-space
domain. (d) Difference between (b) and (c). P and M mark discrepancies on primary and multiple events, respectively.}
\end{figure*}

\begin{figure*}
\includegraphics[width=0.98\textwidth]{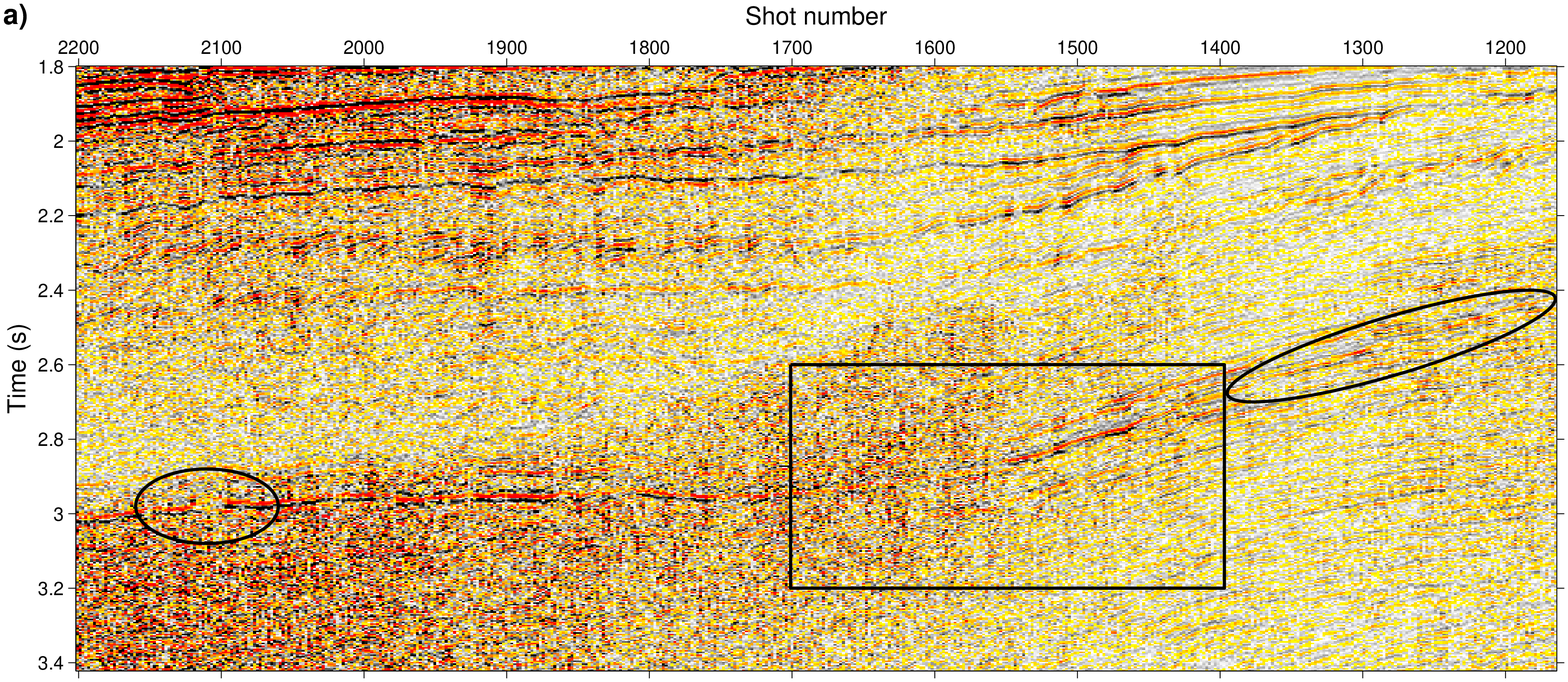}
\includegraphics[width=0.98\textwidth]{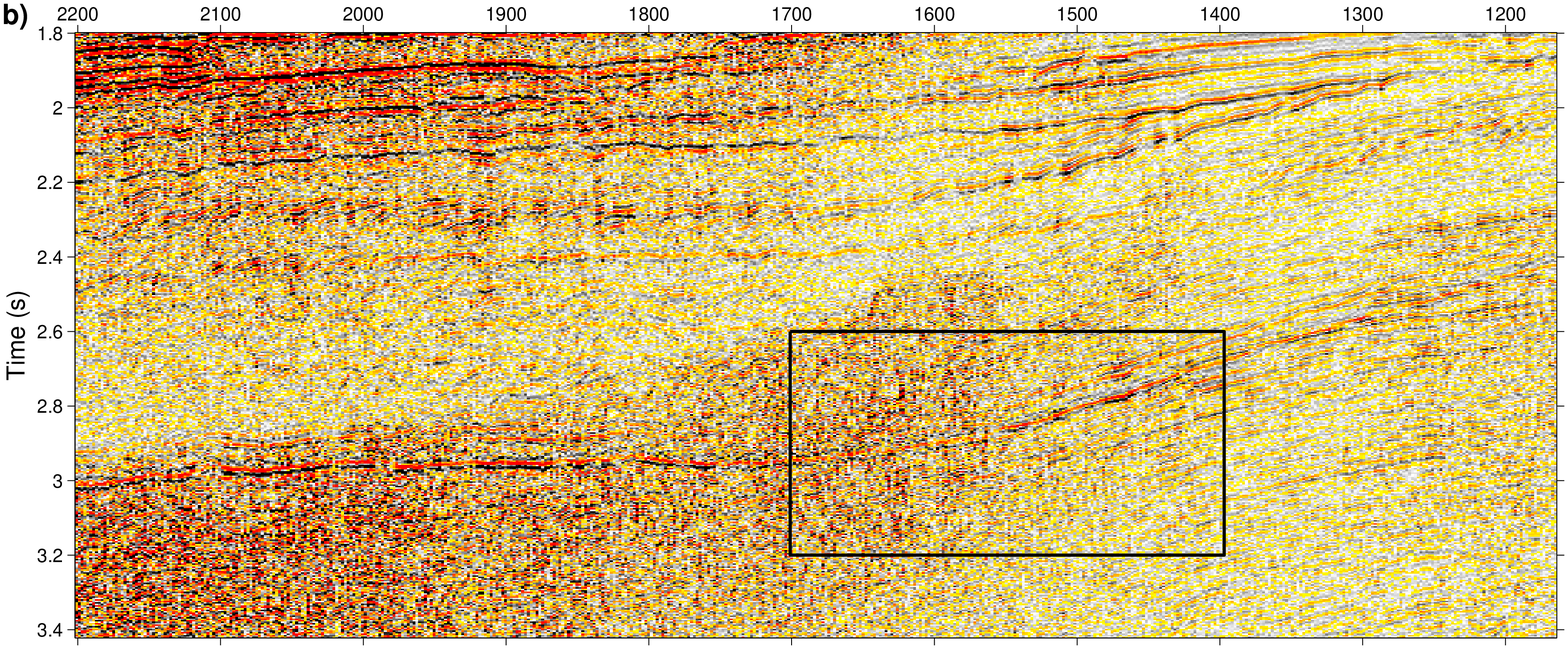}
\includegraphics[width=0.98\textwidth]{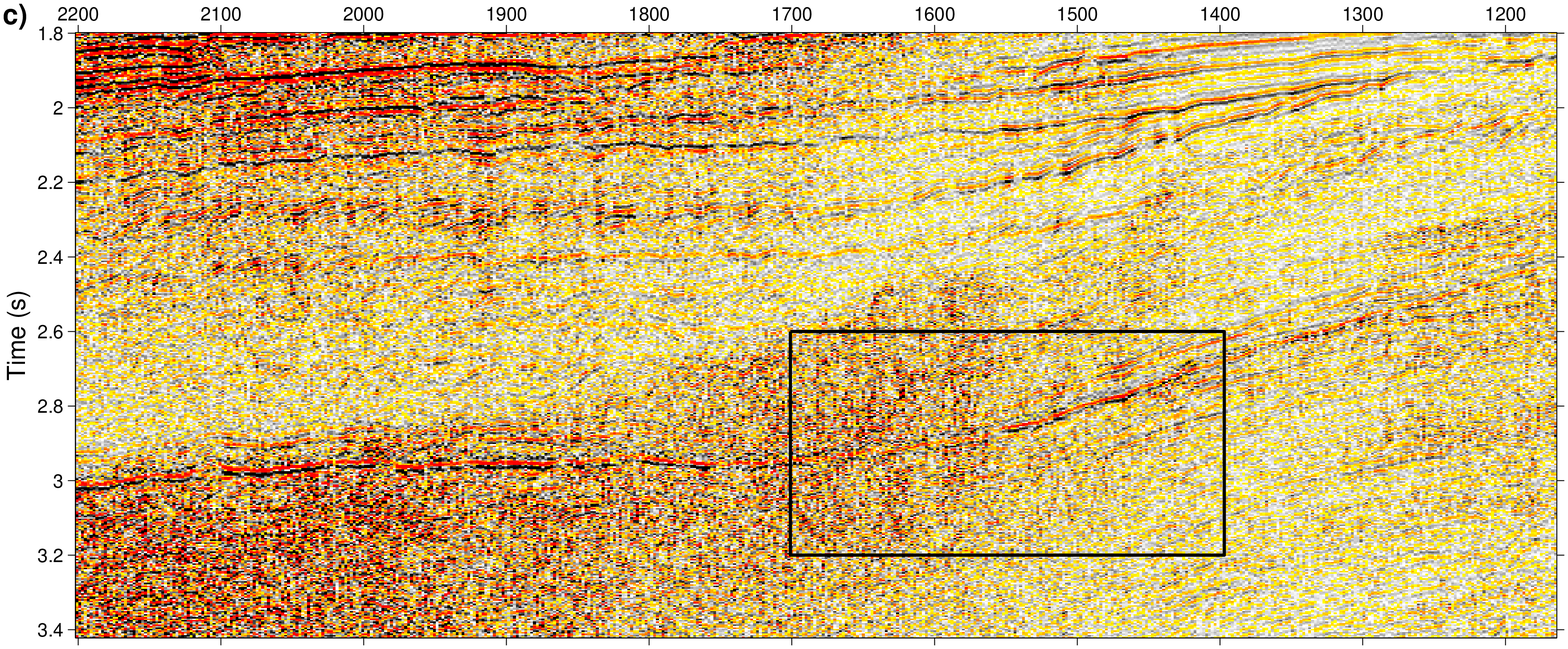}
\caption{\label{fig:DM_FW_CRP1}Subtraction results on the first channel gather.
(a) Standard 1D adaptive filter. (b) 1D  complex unary filters in the time-scale domain. (c) Standard 2D adaptive 
filter in the time-space domain. See Figure \ref{fig:Z3} for a zoom on the delimited zone.}
\end{figure*}

\begin{figure*}
\includegraphics[width=0.49\textwidth]{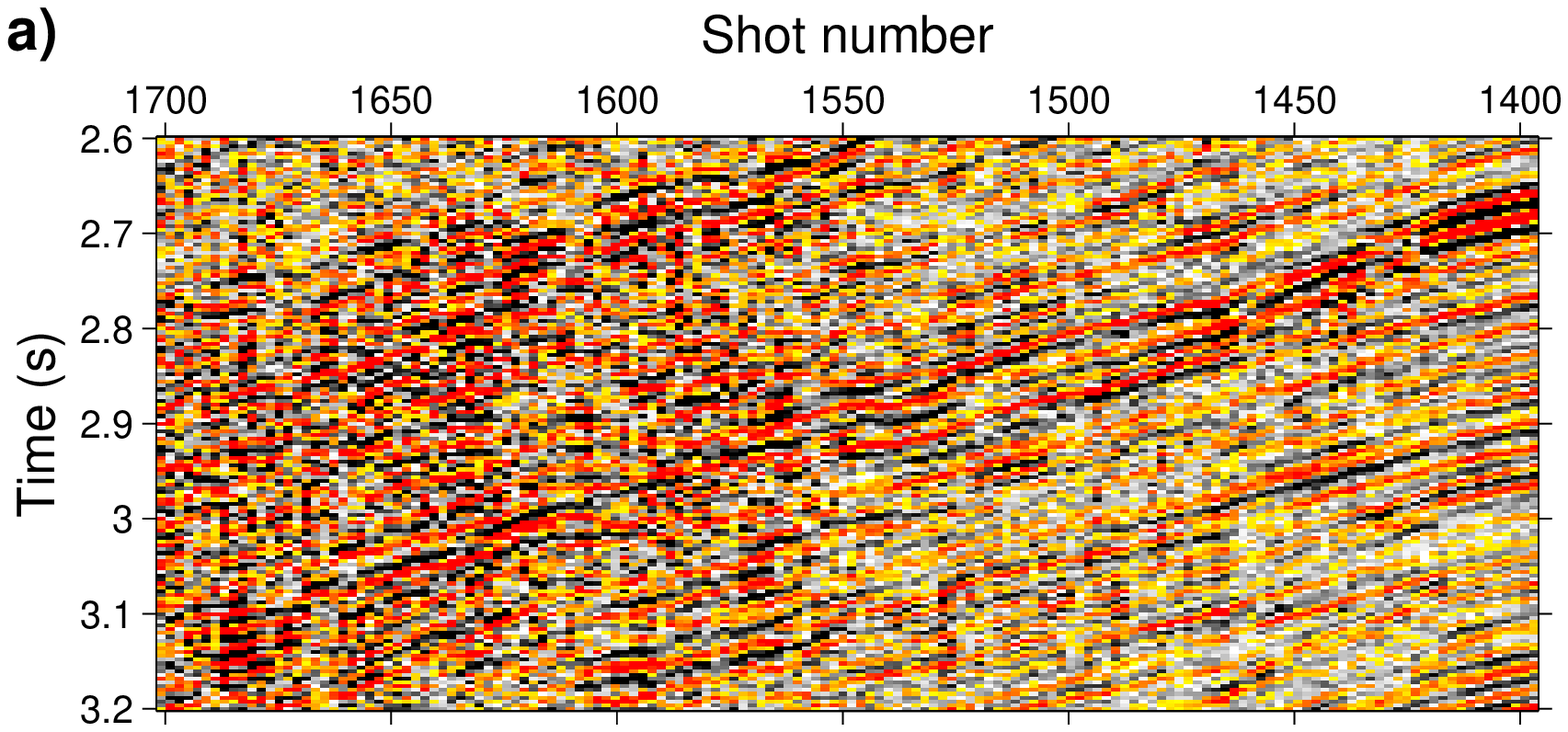}\hfill{}\includegraphics[width=0.49\textwidth]{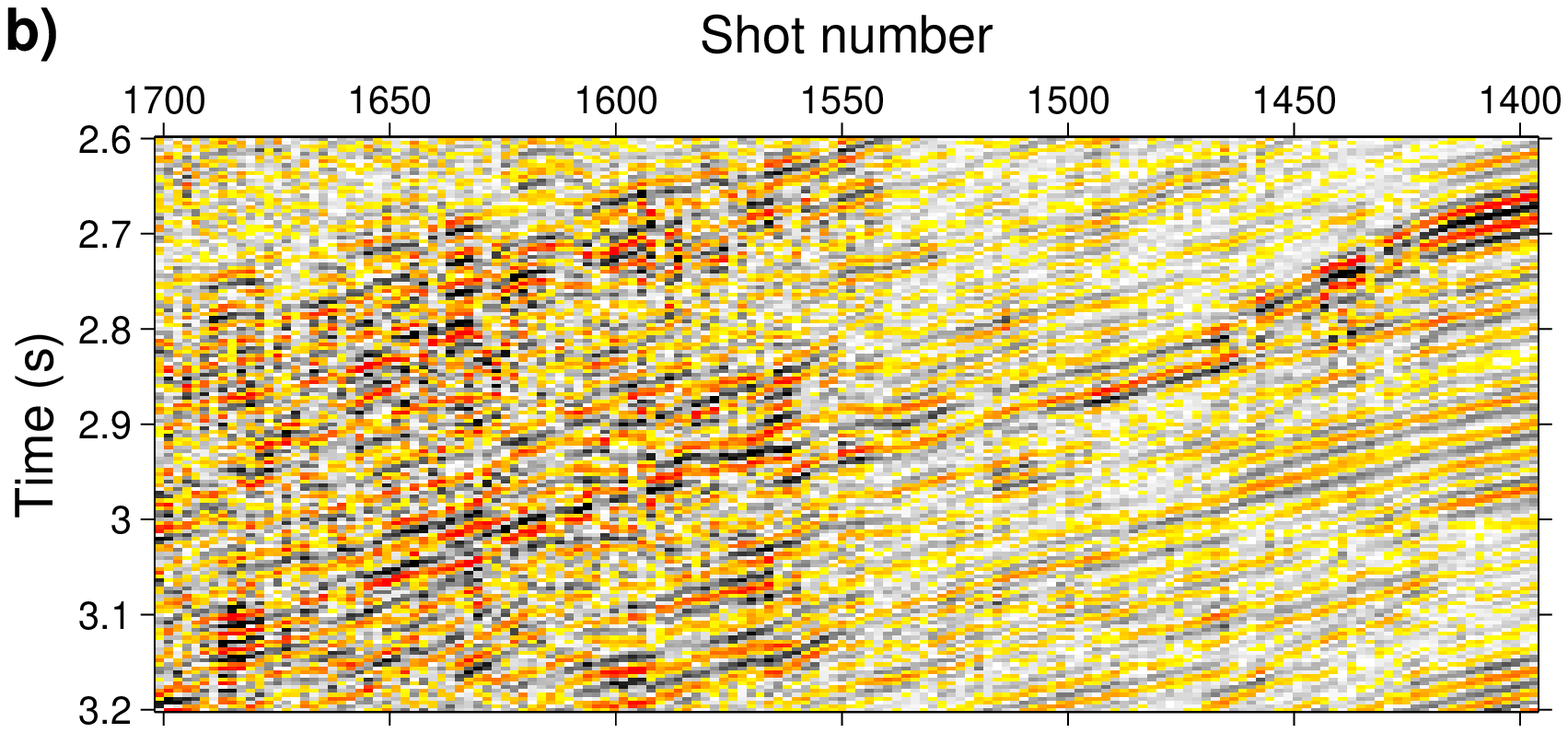}
\includegraphics[width=0.49\textwidth]{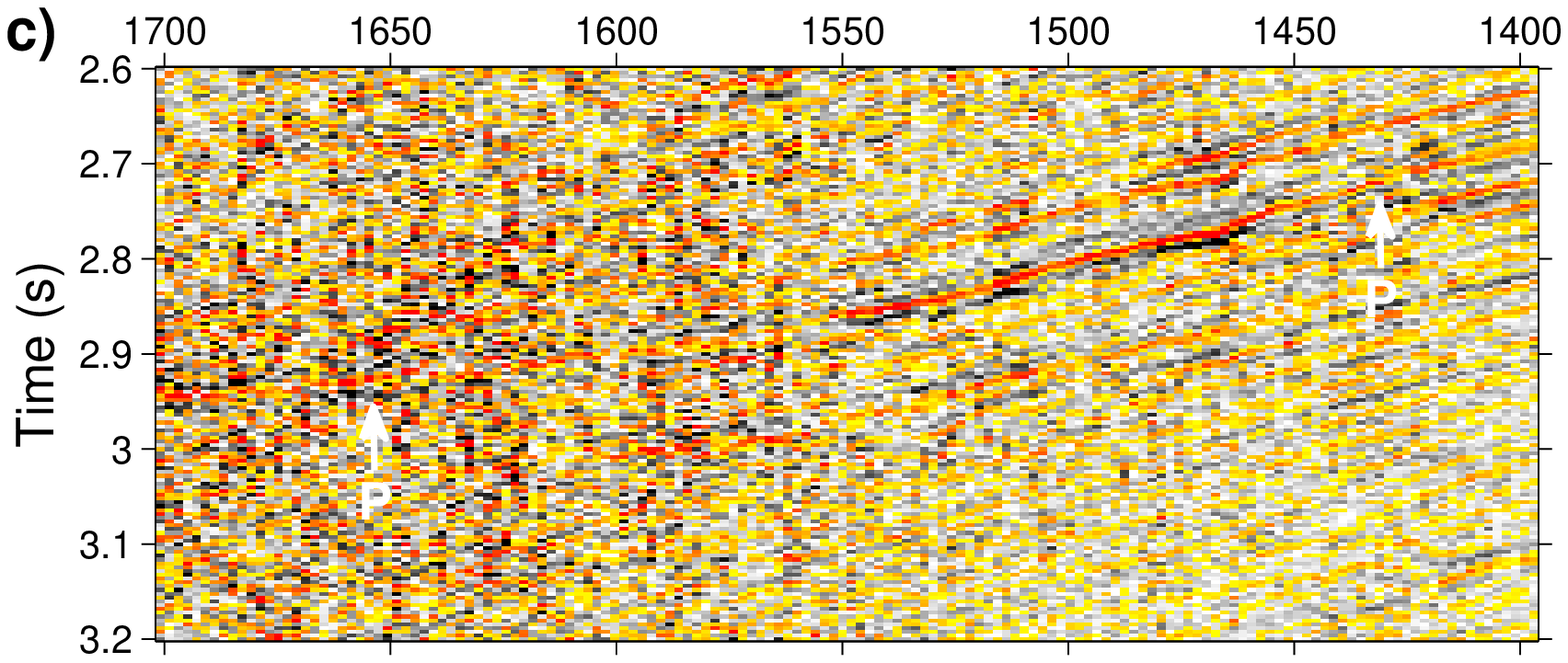}\hfill{}\includegraphics[width=0.49\textwidth]{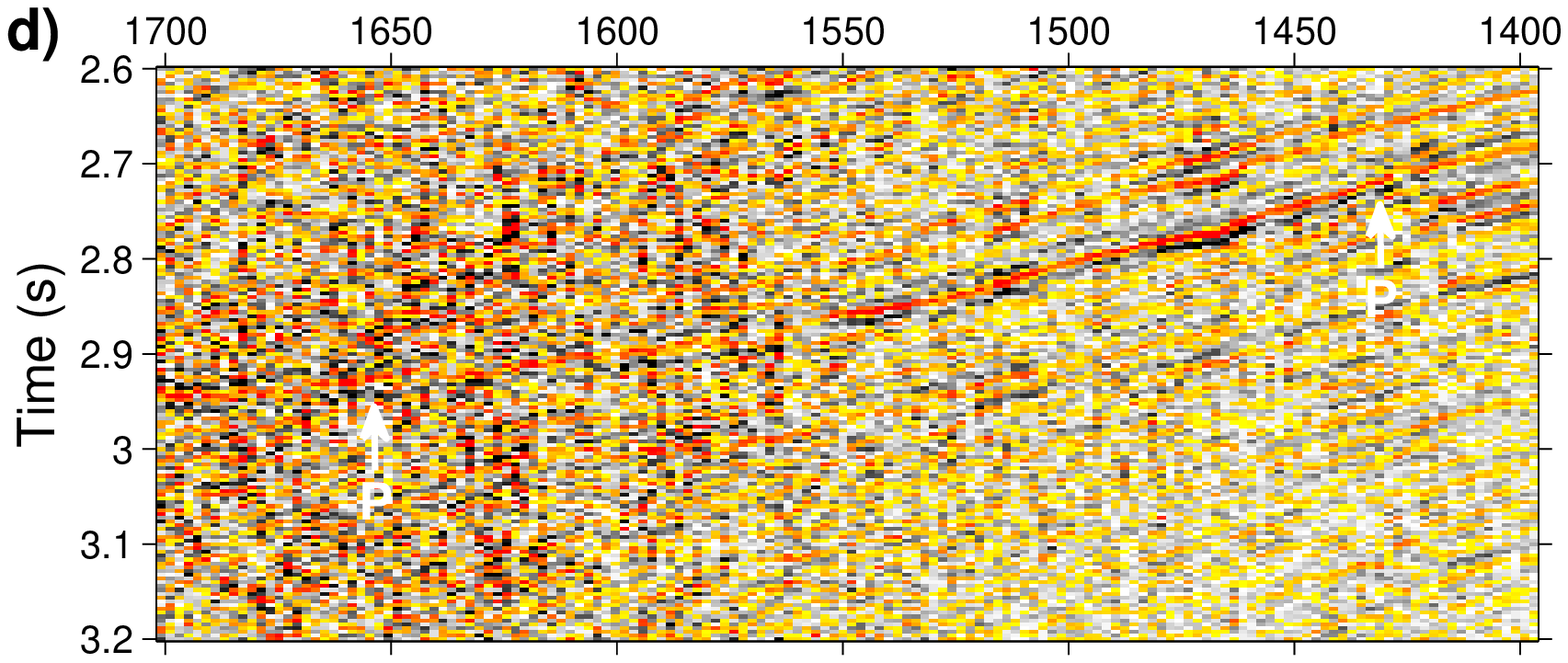}
\includegraphics[width=0.49\textwidth]{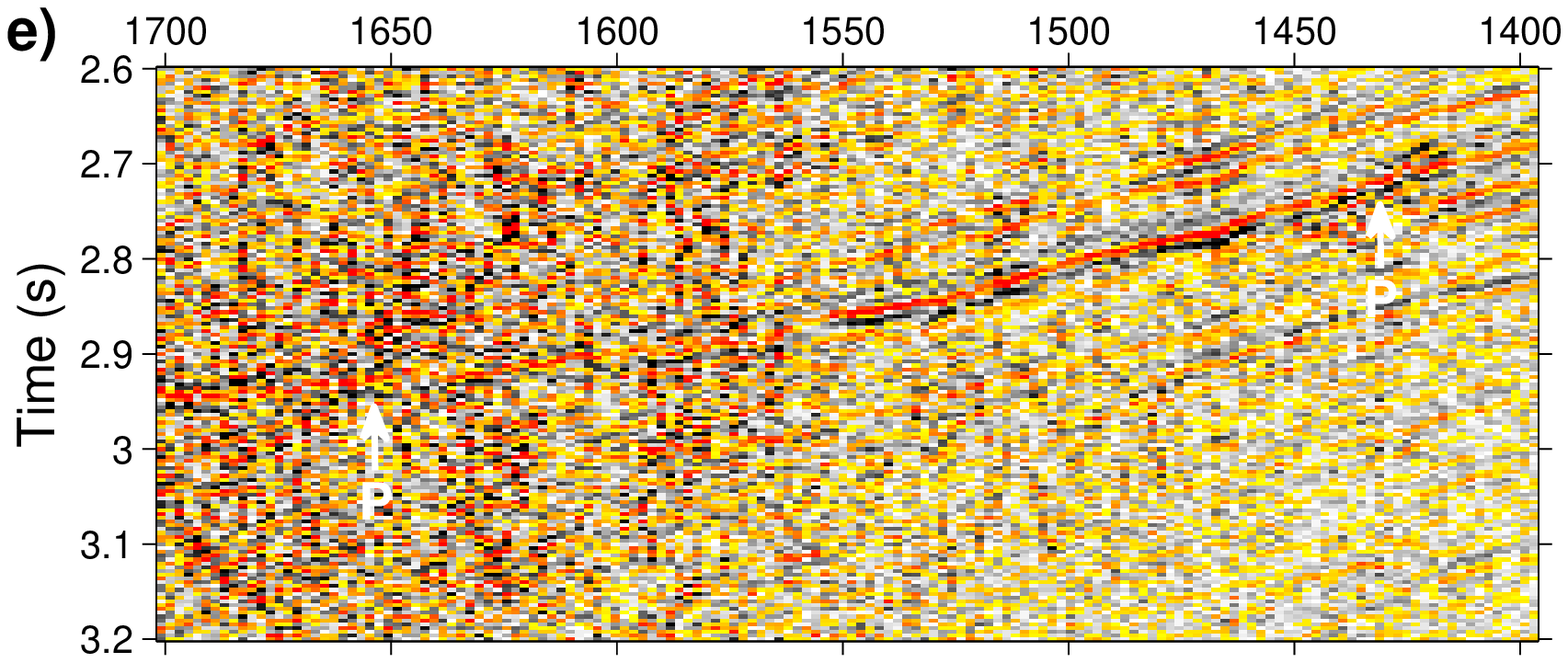}
\caption{\label{fig:Z3} Zoom on traces between shot number 1400 and 1700 of Figure \ref{fig:DM_FW_CRP1}. (a) Recorded data. (b) Original model.  
Filtered data with (c) standard 1D adaptive filters, (d) 1D complex unary filters in the time-scale domain, and (e) standard 2D adaptive filters. P marks discrepancies on primary events.}
\end{figure*}

\begin{figure*}
\includegraphics[width=0.33\textwidth]{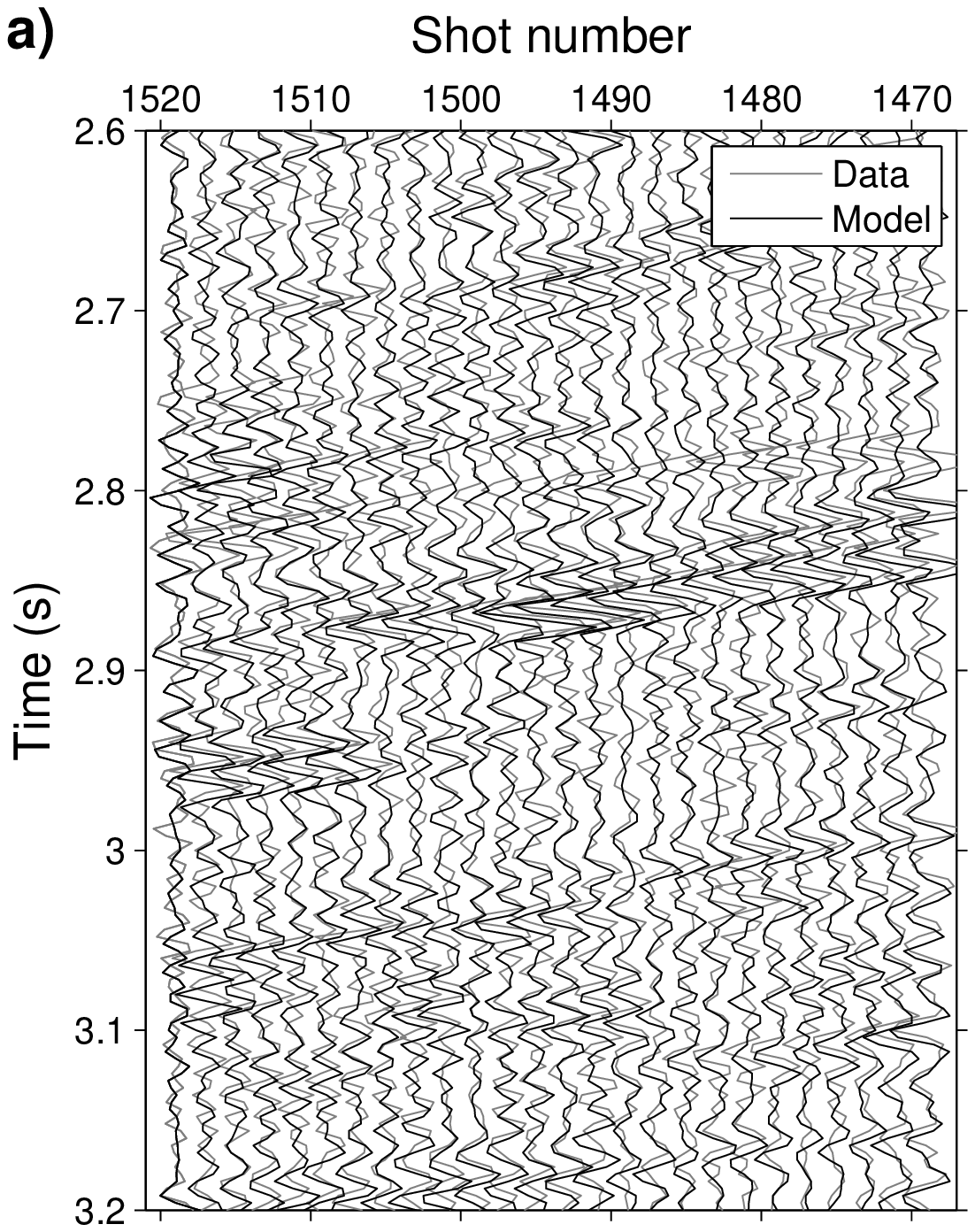}\hfill{}\includegraphics[width=0.33\textwidth]{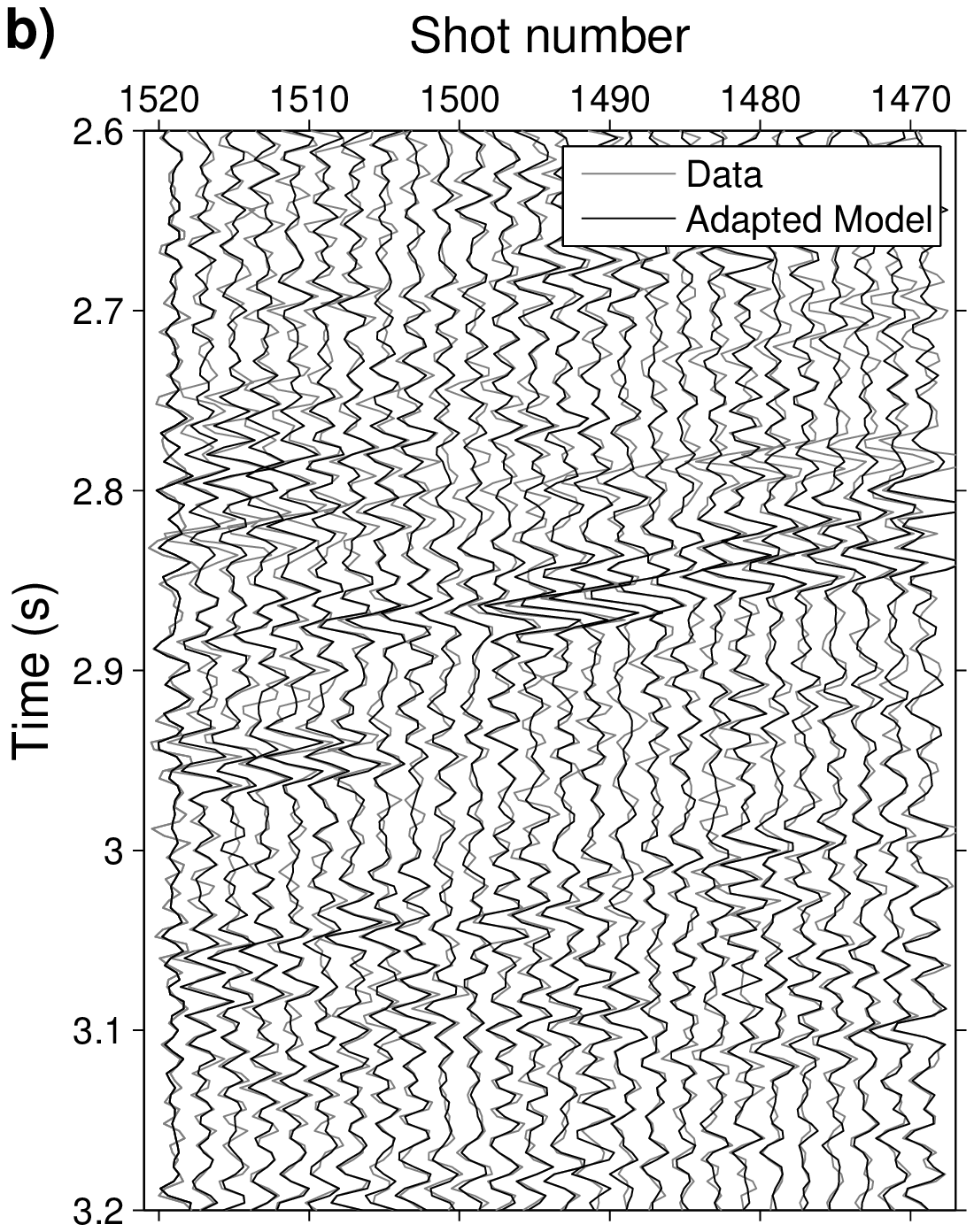}\hfill{}\includegraphics[width=0.33\textwidth]{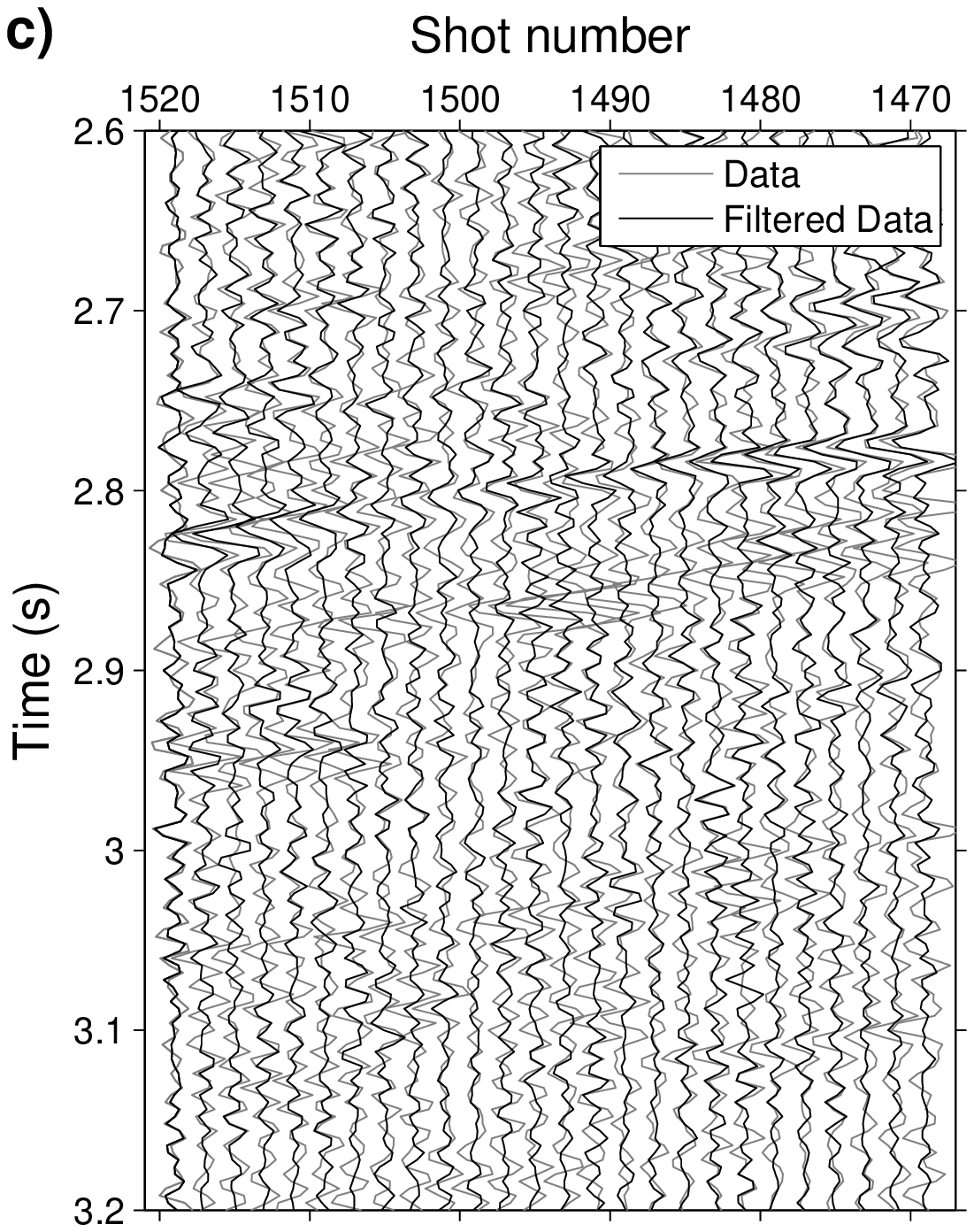}
\caption{\label{fig:Waveforms}Superimposition of selected waveforms. (a) Recorded data and original model with a small positive delay. (b) Recorded data and adapted model with 1D complex unary filters in time-scale. (c) Recorded and filtered data.}
\end{figure*}

\begin{figure*}
\includegraphics[width=0.98\textwidth]{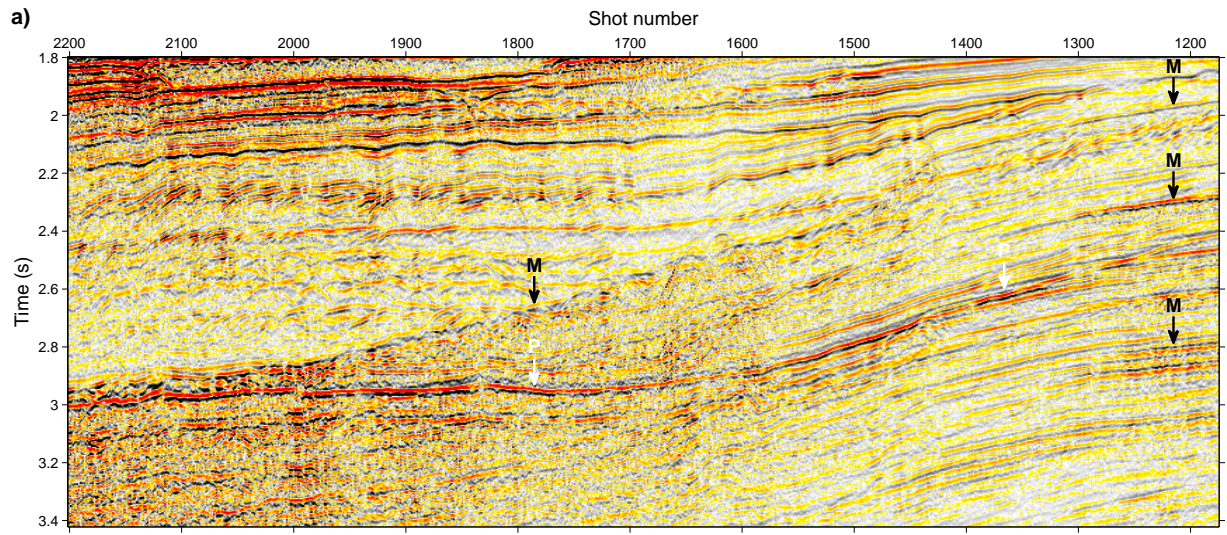}
\caption{\label{fig:STK1}Near offset stack of the raw data. P and M mark primary and multiple events, respectively.}
\end{figure*}

\begin{figure*}
\includegraphics[width=0.98\textwidth]{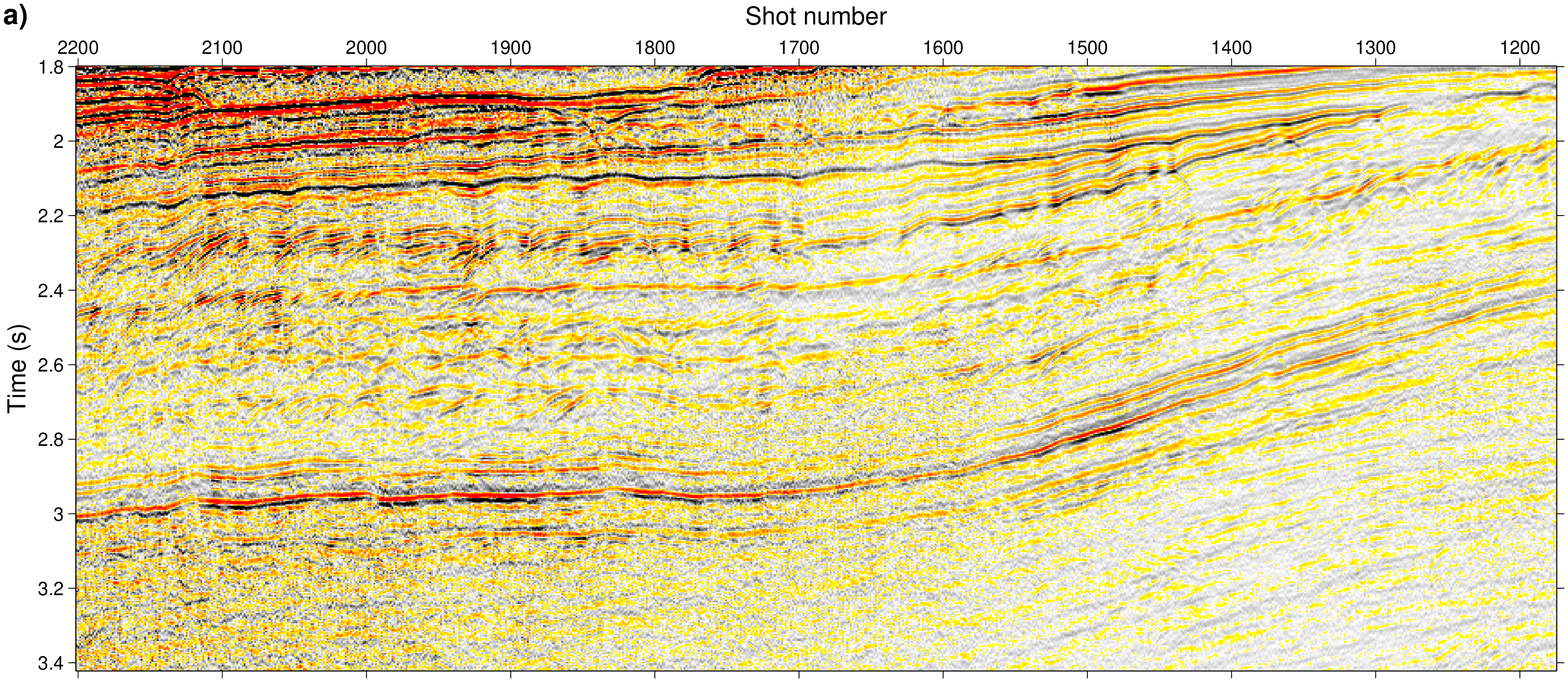}
\includegraphics[width=0.98\textwidth]{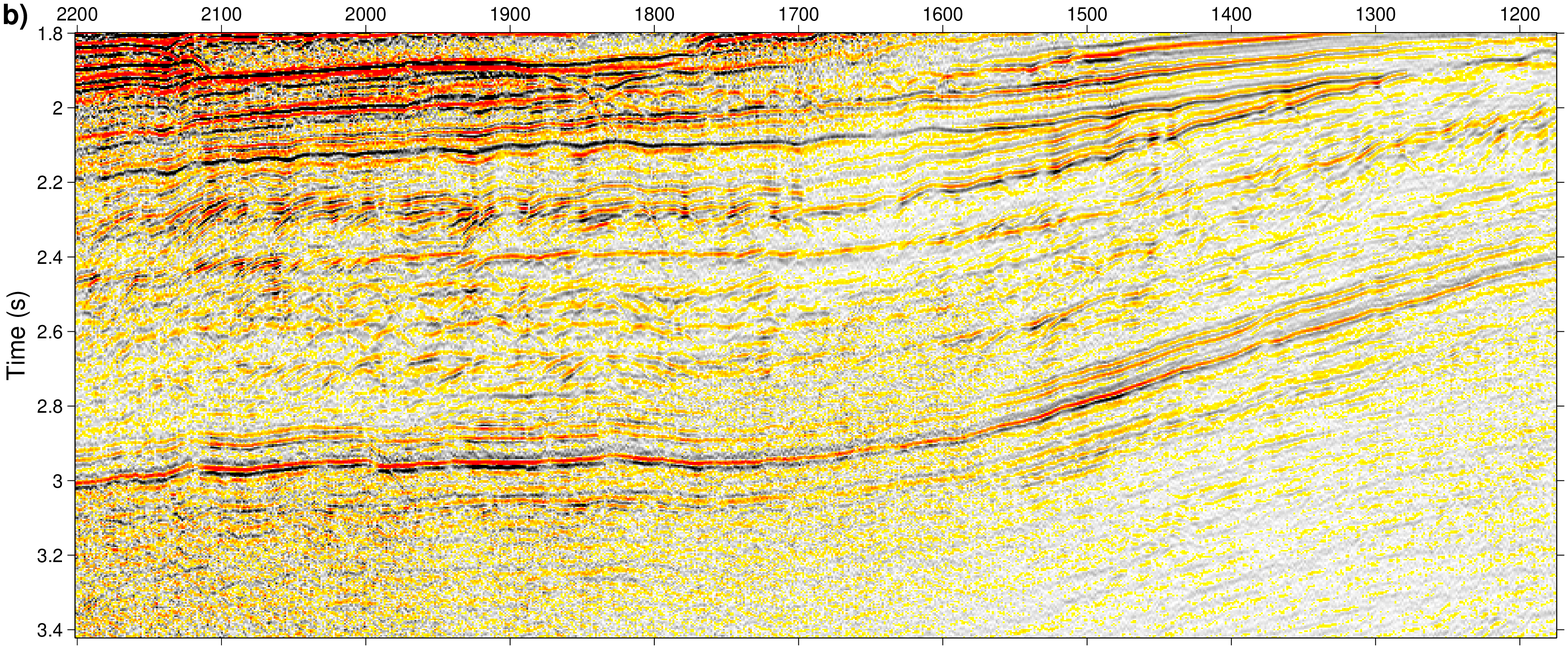}
\includegraphics[width=0.98\textwidth]{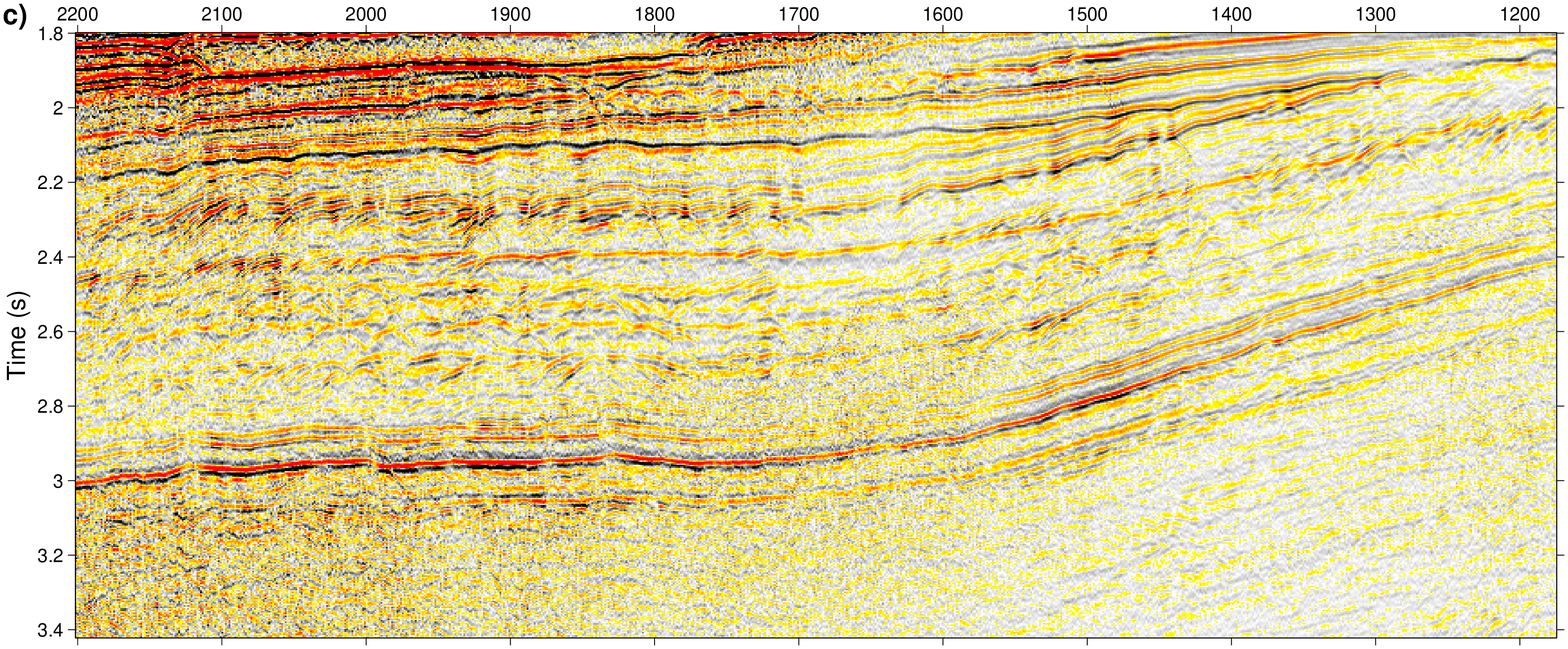}
\caption{\label{fig:STK2}Near offset stacks of filtered data using (a) standard 1D adaptive filters, (b) 1D complex unary filters in time-scale, and (c) standard 2D adaptive filters.}
\end{figure*}

\begin{figure*}
\includegraphics[width=0.98\textwidth]{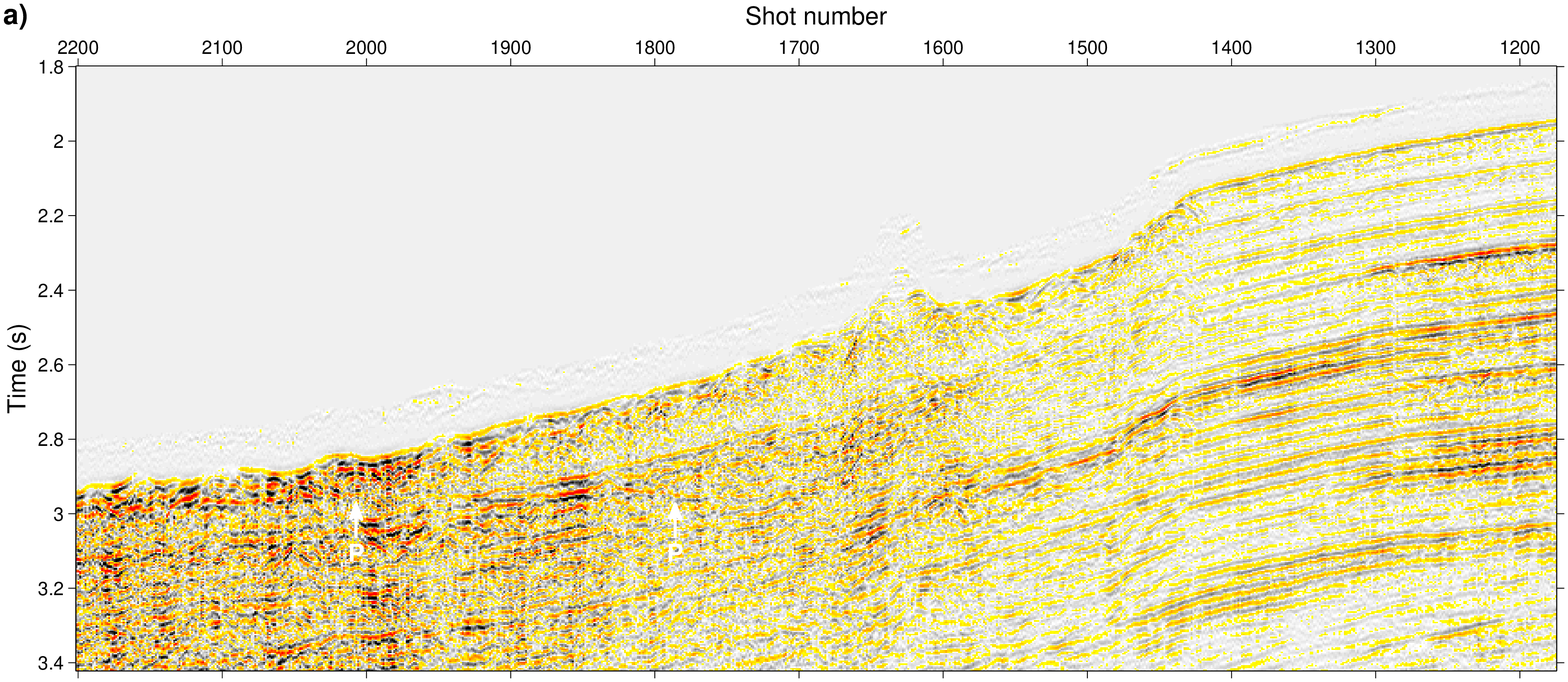}
\includegraphics[width=0.98\textwidth]{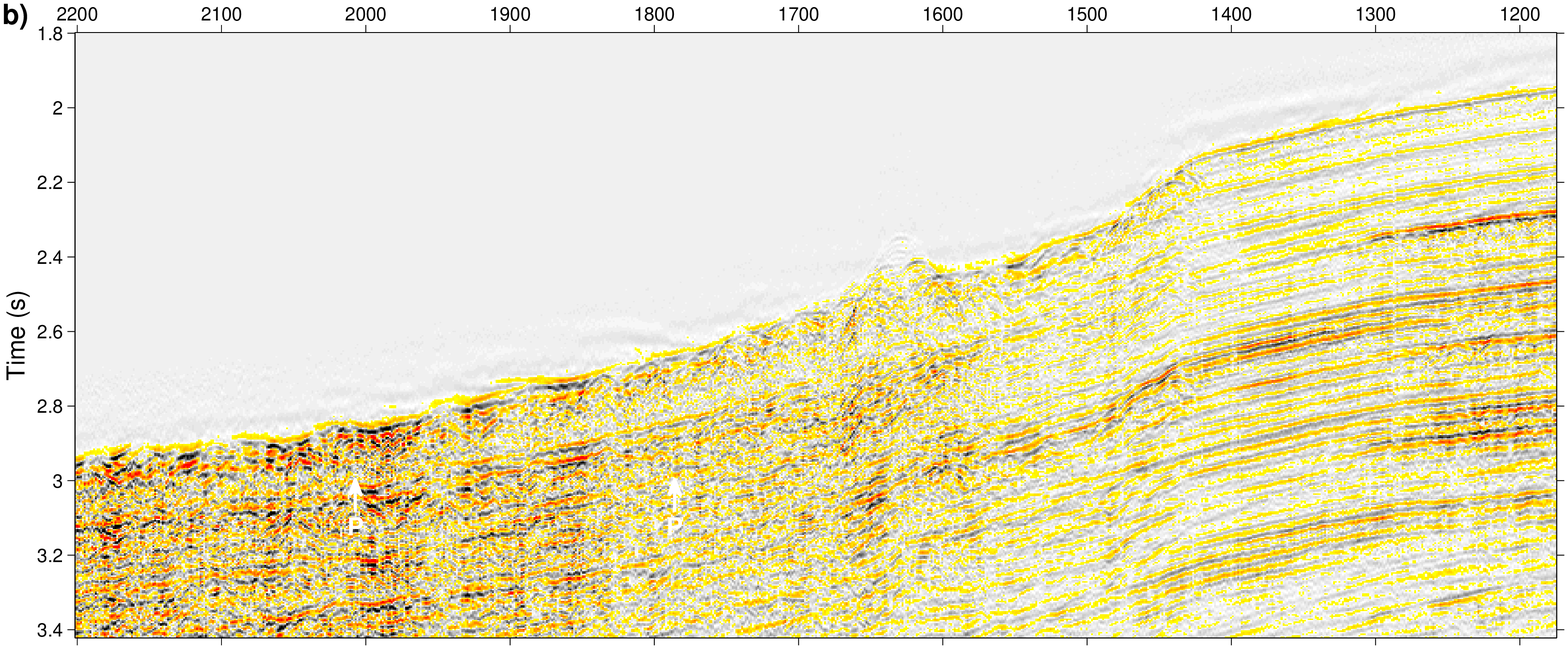}
\includegraphics[width=0.98\textwidth]{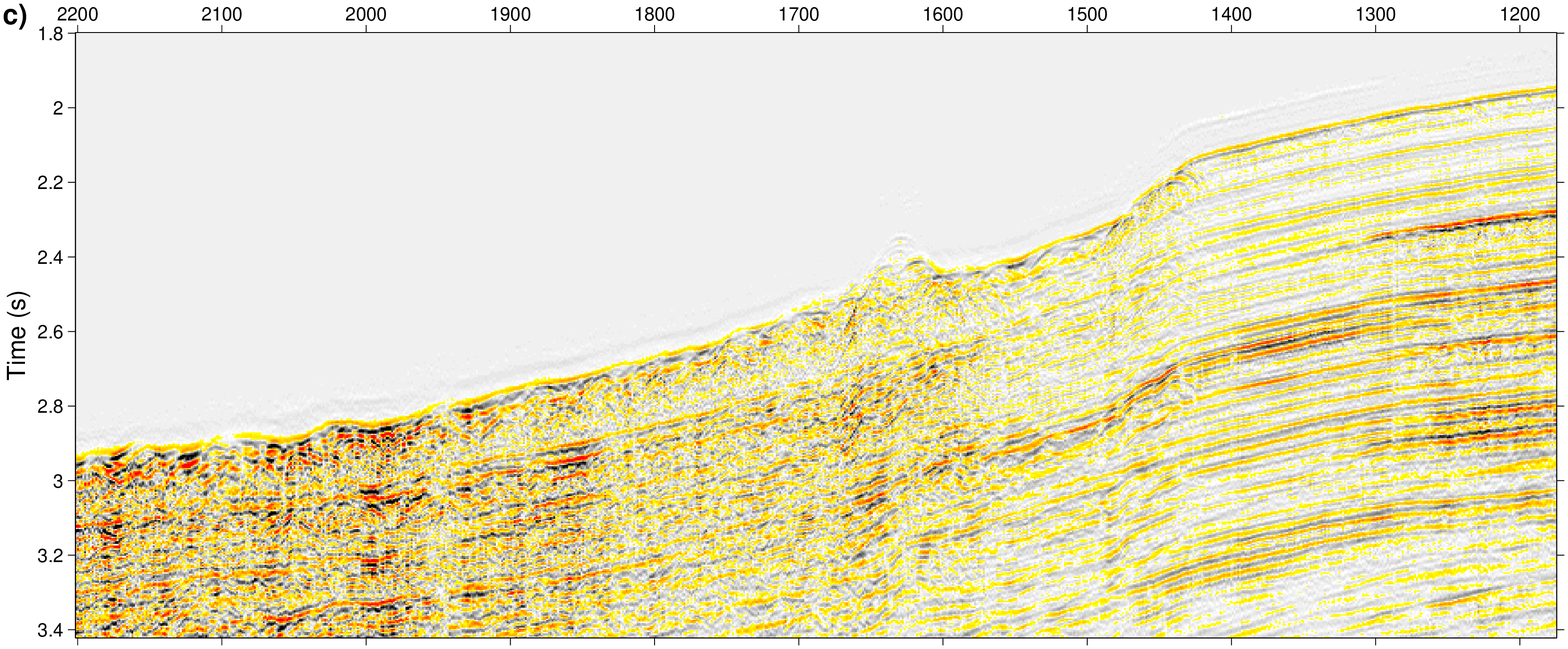}
\caption{\label{fig:STK3}Near offset stacks of the difference between the raw data and the filtered data using (a) standard 1D adaptive filters, (b) 1D complex unary filters in time-scale, and (c) standard 2D adaptive filters. P marks overadaptation to primary events.}
\end{figure*}

Figure \ref{DM_CSP_SYNTH}a represents synthetic data with two intersecting dipping events with wavelets of equal amplitude and frequency content; Figure \ref{DM_CSP_SYNTH}b shows the multiple model. 
As expected, 1D algorithms (1D filters computed in 1D data windows) behave similarly and fail to properly cope with the intersection. 
However, a dataset showing single primary and multiple events with identical wavelet shape is not  completely realistic. 
In practice, variations in wavelet shapes associated  with uncorrelated time arrivals of primaries and multiple  greatly contribute to the success of adaptation in a given time frame.
Thus, better behavior  of the proposed algorithm on actual data is expected. Firstly, it enhances primaries and multiples discrepancies in the wavelet domain. Secondly, it enables  cross-correlation reduction using larger window frames containing several signals, as demonstrated in the following field example.

The examples shown in Figures \ref{fig:DM_CSP}a and \ref{fig:DM_CRP1}a
are taken from a 3D real marine dataset, in common shot and channel gather, respectively.
The main objective is to uncover primaries masked by strong multiple events using a multiple model. The 3D-SRME model \citep{Pica_A_2005_j-tle_3d_srmm} shown in Figures \ref{fig:DM_CSP}b and \ref{fig:DM_CRP1}b  achieves good precision on the near and mid offset range. 
We compare the proposed method based on 1D  complex unary filters in wavelet domain, with  reference algorithms based on standard 1D and 2D adaptive filters.
Figures \ref{fig:DM_FW_FA_CSP} and \ref{fig:DM_FW_CRP1} represent subtraction results in common shot and channel gathers, respectively. 
Additional details are provided in Figures \ref{fig:Z3} (for improved visual amplitude assessment) and \ref{fig:Waveforms} (for accurate waveform comparison), to better illustrate the main features in data and differences between algorithms.

With a standard two-step method, the global step mainly contains waveform and time-shift corrections, combined with a global amplitude adaptation factor.
In 2D, it is obtained with large 2D windows and long 2D filters. 
The global step is followed by a local adaptation,  to account for local amplitude and phase discrepancies (in 2D, with small 2D windows and short 1D filters). 
Reference 2D algorithm parameters are taken from the best result obtained in 
an independent benchmark. Despite not being used in practice, the 1D adaptive filter is provided for completeness only.

With the proposed one-step method, a Morlet wavelet frame (with parameters $\omega_{0} = 6.4$ rad/s, $j\in[1,4]$, $v\in[0,3]$ and $b_{0}=1$) is used to decompose the data and reconstruct the filtered signal. 
In the wavelet transformed domain, the optimum unary filter estimation for each sample uses a rectangular window of $0.636$ s.
Additionally, to provide robustness against delays higher than half the signal period (at each scale), we have chosen the maximum normalized cross-correlation criterion within a range of $\pm12$ ms. This subtraction operator 
may compensate for additional distortions as well, 
e.g., wavelet squaring introduced in convolution-based multiple models, since the estimation of the inverse wavelet
is implicit in the estimation of an LSE filter.

Being single-pass, wavelet-based unary Wiener filters achieve slightly better non-coherent noise level (e.g., in Figure \ref{fig:Z3}). 
As pointed out in ellipses in Figure \ref{fig:DM_FW_CRP1}, standard 1D lacks  in terms of primary preservation and multiple attenuation. While the standard 2D approach exhibits best performance to this respect, as shown in the zoom of Figure \ref{fig:DM_FW_CRP1}, the proposed method exhibits close performance, despite not using additional information conveyed by neighboring traces.
Figure \ref{fig:Waveforms} compares the waveforms of data, model, adapted and filtered results, for traces between indices 1470 and 1520 from Figure \ref{fig:Z3}. 
We appreciate how the set of unary filters successfully reduces the differences between  multiple model and actual multiple events, to attain sufficient multiple event attenuation to uncover main primary events. The trace overlay indicate the waveforms are well preserved. 

To complete the picture, we provide  additional results on near offset stacks. 
 Figures \ref{fig:STK1} to \ref{fig:STK3}  compare 
raw data and the different adaptive multiple subtraction methods.
Note that stacking  removes a large portion of the multiples. The performance is thus evaluated on finer details, especially in the difference between raw data and filtered sections, as shown in Figure \ref{fig:STK3}.
The proposed 1D  wavelet-based complex unary filter appears less aggressive with respect to the primaries previously masked by the multiples than the global adaptation step of a standard 1D method (after the local step, over-adaptation was even higher), compare the pointed signals in Figures \ref{fig:STK3}a and \ref{fig:STK3}b.
However, as previously inferred, the proposed method leaves a little more remnant multiples than the standard 2D method, as it can be appreciated by comparing figures around shot number 1200 and above.

The continuous wavelet frame provides an inherent scale-adaptive windowing
that, when combined with the 1D complex unary filter, offers 
obvious improvements over standard 1D LSE time-varying frequency filters,
and compares promisingly, at least away from isolated event crossings, with the industry standard 
2D method on the common-channel gather, 
where a perfect removal of the source effects can still be 
demanding. Additional constraints may be added in the scale and in
the space dimensions to improve the fractional
delay estimation and reinforce the lateral coherence of the method, either with adaptive \citep{Donno_D_2011_j-geophysics_improving_mrulsdfica,Ventosa_S_2012_j-geophysics_window_lsosrlsst} or 
low-redundancy multiscale transforms \citep{Chaux_C_2007_j-ieee-tit_noi_cpdtwd}, which are robust to correlated noises and possess finer bandwidth selectivity than dyadic scales.

\section{Conclusion}

We propose a model-based multiple subtraction which combines complex Morlet 
wavelet frame with  complex unary Wiener filters. This subtraction method 
counter-balances the redundancy provided by this complex wavelet frame with 
the freedom of designing fast filters which allow for simple non-stationary 
model adaptation. Complex unary filters with a low varying phase along scale 
can adapt fractional delays up to half the signal period at a given frequency. 
To add robustness against higher delays, we integrate an integer delay term in 
the kernel of the unary filters, which can be estimated globally or locally, 
depending on the application. Despite not competing with more advanced multiple 
removal methods, either regularized or in higher dimension spaces, the results 
of 1D unary filters in time-scale compare promisingly with standard 2D LSE 
methods. The structure of the algorithm allows for reduced memory footprint, 
higher code parallelization, and is potentially suitable to sparse acquisition. 
Further developments foresee additional diversity borrowed from neighboring 
traces, to improve apparent velocity separation and residual noise attenuation.
The presented multiple removal technique aims at updating the portfolio of 
hybrid demultiple algorithms.

\section{Acknowledgments}
The authors thank Statoil for allowing them to show the Norwegian Sea results. They also acknowledge  IFP Energies nouvelles and CGGVeritas for the authorization to present this work. They are indebted to Anatoly Baumstein, Ramesh Neelamani, Tamas Nemeth and the anonymous reviewers for their constructive comments that helped improve significant parts of the paper. They finally recognize Bruno Lety (IFPEN) and Nicolas Quash  (CGGVeritas) for their careful proofreading.

\bibliographystyle{alpha}

\begin{thebibliography}{}
\itemsep0pt

\bibitem[Abma et~al., 2005]{Abma_R_2005_j-tle_com_asmma}
Abma, R., N. Kabir, K.~H. Matson, S. Michell, and S. A.  Shaw and B.~McLain,
  2005, Comparisons of adaptive subtraction methods for multiple attenuation:
  The Leading Edge, {\bf 24}, 277--280.

\bibitem[Ahmed, 2007]{Ahmed_I_2007_p-seg_2d_wtdase3dsrme}
Ahmed, I.,  2007, {2D} wavelet transform-domain adaptive subtraction for
  enhancing {3D} {SRME}: 77th Annual International Meeting, SEG, Expanded 
  Abstracts, 2490--2494.
  
  \bibitem[Amundsen et~al.,
  2001]{Amundsen_L_2001_j-geophysics_multidimensional_sdfsmemmobsd}
Amundsen, L., L.~T. Ikelle, and L.~E. Berg,  2001, Multidimensional signature
  deconvolution and free-surface multiple elimination of marine multicomponent
  ocean-bottom seismic data: Geophysics, {\bf 66}, 1594--1604.

\bibitem[Berkhout and Verschuur,
  2006]{Berkhout_A_2006_j-geophysics_foc_ticsrnr}
Berkhout, A.~J., and D.~J. Verschuur,  2006, Focal transformation, an imaging
  concept for signal restoration and noise removal: Geophysics, {\bf 71},
  A55--A59.

\bibitem[Buttkus, 1975]{Buttkus_B_1975_j-geophys-prospect_hom_ftp}
Buttkus, B.,  1975, Homomorphic filtering --- theory and practice: Geophysical
  Prospecting, {\bf 23}, 712--748.
  
  \bibitem[Chaux et~al., 2007]{Chaux_C_2007_j-ieee-tit_noi_cpdtwd}
Chaux, C., J.-C. Pesquet, and L. Duval,  2007, Noise covariance properties in
  dual-tree wavelet decompositions: IEEE Transactions on  Information Theory, {\bf 53},
  4680--4700.

\bibitem[de~Hoop et~al., 2009]{DeHoop_M_2009_j-inv-prob_sei_igrtctp}
de~Hoop, M.~V., H. Smith, G. Uhlmann, and R.~D. van~der Hilst,  2009, Seismic
  imaging with the generalized {Radon} transform: a curvelet transform
  perspective: Inverse Problems, {\bf 25}, 025005 (21pp).

\bibitem[Donno et~al., 2010]{Donno_D_2010_j-geophysics_cur_bmp}
Donno, D., H. Chauris, and M. Noble,  2010, Curvelet-based multiple prediction:
  Geophysics, {\bf 75}, WB255--WB263.
  
  \bibitem[Donno, 2011]{Donno_D_2011_j-geophysics_improving_mrulsdfica}
Donno, D., 2011, Improving multiple removal using least-squares dip filters and independent
 component analysis:
  Geophysics, {\bf 76}, V91--V104.

\bibitem[Dragoset et~al., 2010]{Dragoset_B_2010_j-geophysics_per_3dsrme}
Dragoset, B., E. Verschuur, I. Moore, and R. Bisley,  2010, A perspective on
  {3D} surface-related multiple elimination: Geophysics, {\bf 75},
  75A245--75A261.

\bibitem[Fomel, 2009]{Fomel_S_2009_j-geophysics_ada_msrnr}
Fomel, S.,  2009, Adaptive multiple subtraction using regularized nonstationary
  regression: Geophysics, {\bf 74}, V25--V33.

\bibitem[Gilloire and Vetterli,
  1992]{Gilloire_V_1992_j-ieee-tsp_ada_fsbcsaeaaec}
Gilloire, A., and M. Vetterli,  1992, Adaptive filtering in sub-bands with
  critical sampling: analysis, experiments, and application to acoustic echo
  cancellation: IEEE Transactions on Signal Processing, {\bf 40}, 1862--1875.

\bibitem[Guitton, 2004]{Guitton_A_2004_j-geophys-prospect_pat_bamra3dgmds}
Guitton, A.,  2004, A pattern-based approach for multiple removal applied to a
  {3D} {G}ulf of {M}exico data set: Geophysical Prospecting, {\bf 54},
  135--152.

\bibitem[Guitton and Verschuur,
  2004]{Guitton_A_2004_j-geophys-prospect_ada_smul1n}
Guitton, A., and D.~J. Verschuur,  2004, Adaptive subtraction of multiples
  using the $l_1$-norm: Geophysical Prospecting, {\bf 52}, 27--38.

\bibitem[Hampson, 1986]{Hampson_D_1986_j-can-j-explor-geophys_inverse_vsme}
Hampson, D.,  1986, Inverse velocity stacking for multiple elimination: Canadian Journal of Exploration Geophysicists, {\bf 22}, 44--55.

\bibitem[Herrmann and Verschuur, 2005]{Herrmann_F_2005_p-eage_rob_cdpmssc}
Herrmann, F., and D.~J. Verschuur, 2005, Robust curvelet-domain primary-multiple separation with
  sparseness constraints: 67th Annual Meeting, EAGE, Expanded Abstracts, 
  P226.
  
\bibitem[Herrmann et~al., 2000]{Herrmann_P_2000_p-seg_de_ahrrt}
Herrmann, P., T. Mojesky, T. Magesan, and P. Hugonnet,  2000, De-aliased,
  high-resolution {R}adon transforms: 70th Annual International Meeting, 
  SEG, Expanded Abstracts, 1953--1956.

\bibitem[Huo and Wang, 2009]{Huo_S_2009_j-geophysics_imp_assma}
Huo, S., and Y. Wang,  2009, Improving adaptive subtraction in seismic multiple
  attenuation: Geophysics, {\bf 74}, 59--67.

\bibitem[Jacques et~al., 2011]{Jacques_L_2011_j-sp_pan_mgrisdfs}
Jacques, L., L. Duval, C. Chaux, and G. Peyr{\'e},  2011, A panorama on
  multiscale geometric representations, intertwining spatial, directional and
  frequency selectivity: Signal Processing, {\bf 91}, 2699--2730.

\bibitem[Jorgensen and Song, 2009]{Jorgensen_P_2009_incoll_com_dcwt}
Jorgensen, P. E.~T., and M.-S. Song,  2009, Comparison of discrete and
  continuous wavelet transforms, {\it in} Encyclopedia of Complexity and
  Systems Science: Springer, {\bf 1-10}.

\bibitem[Kabir and Verschuur, 2007]{Kabir_N_2007_j-cseg-recorder_par_rtmrhaavoa}
Kabir, N., and D.~J. Verschuur,  2007, Expert answers: Does parabolic Radon
  transform multiple removal hurt amplitudes for AVO analysis?: CSEG Recorder,
  10--14.

\bibitem[Kaplan and Innanen, 2008]{Kaplan_S_2008_j-geophysics_ada_sfsmica}
Kaplan, S.~T., and K.~A. Innanen,  2008, Adaptive separation of free-surface
  multiples through independent component analysis: Geophysics, {\bf 73},
  V29--V36.
    
\bibitem[Lin et~al., 2004]{Lin_D_2004_p-seg_3d_srmeagm}
Lin, D., J. Young, Y.  Huang, M. Hartmann,  2004, {3D} {SRME} application in the {G}ulf of {M}exico: 74th Annual International Meeting, SEG, Expanded 
  Abstracts, 1257--1260.

\bibitem[Lines, 1996]{Lines_L_1996_j-can-j-explor-geophys_sup_spmdmbi}
Lines, L.,  1996, Suppression of short-period multiples --- deconvolution or
  model-based inversion?: Canadian Journal of Exploration Geophysics, 
  {\bf 32}, 63--72.

\bibitem[Monk, 1993]{Monk_D_1993_j-geophys-prospect_wav_emscge}
Monk, D.~J.,  1993, Wave-equation multiple suppression using constrained
  cross-equalization: Geophysical Prospecting, {\bf 41}, 725--736.

\bibitem[Neelamani et~al., 2010]{Neelamani_R_2010_j-geophysics_ada_scvct}
Neelamani, R., A. Baumstein, and W.~S. Ross,  2010, Adaptive subtraction
  using complex-valued curvelet transforms: Geophysics, {\bf 75}, V51--V60.

\bibitem[Neidell and Taner, 1971]{Neidell_N_1971_j-geophysics_sem_ocmmd}
Neidell, N.~S., and M.~T. Taner,  1971, Semblance and other coherency measures
  for multichannel data: Geophysics, {\bf 36}, 482--497.

\bibitem[Nowak and Imhof, 2006]{Nowak_E_2006_j-geophysics_amp_prbmrf}
Nowak, E.~J., and M.~G. Imhof,  2006, Amplitude preservation of {Radon}-based
  multiple-removal filters: Geophysics, {\bf 71}, V123--V126.

\bibitem[Nuzzo and Quarta, 2004]{Nuzzo_L_2004_j-geophysics_imp_gprcnatpwt}
Nuzzo, L., and T. Quarta,  2004, Improvement in {GPR} coherent noise
  attenuation using $\tau-p$ and wavelet transforms: Geophysics, {\bf 69},
  789--802.

\bibitem[Pesquet and Leporini, 1997]{Pesquet_J_1997_icipa_new_weid}
Pesquet, J.-C., and D. Leporini,  1997, A new wavelet estimator for image
  denoising: IEE Sixth International Conference on Image Processing and Its Applications, 249--253.

\bibitem[Pica et~al., 2005]{Pica_A_2005_j-tle_3d_srmm}
Pica, A., G. Poulain, B. David, M. Magesan, S. Baldock, T. Weisser, P.
  Hugonnet, and P. Herrmann,  2005, {3D} surface-related multiple modeling: The
  Leading Edge, {\bf 24}, 292--296.

\bibitem[Pokrovskaia and Wombell, 2004]{Pokrovskaia_T_2004_p-eage_att_rmcnwtd}
Pokrovskaia, T., and R. Wombell,  2004, Attenuation of residual multiples and
  coherent noise in the wavelet transform domain: 66th Annual Meeting, 
  EAGE, Expanded Abstracts, P175.

\bibitem[Ristow and Kosbahn, 1979]{Ristow_D_1979_j-geophys-prospect_tim_vpfmu}
Ristow, D., and B. Kosbahn,  1979, Time-varying prediction filtering by means
  of updating: Geophysical Prospecting, {\bf 27}, 40--61.

\bibitem[Robinson and Treitel,
  1967]{Robinson_E_1967_j-geophys-prospect_pri_dwf}
Robinson, E.~A., and S. Treitel,  1967, Principles of digital {W}iener
  filtering: Geophysical Prospecting, {\bf 15}, 311--332.

\bibitem[Schimmel and Paulssen,
  1997]{Schimmel_M_1997_j-geophys-j-int_noi_rdwcstpws}
Schimmel, M., and H. Paulssen,  1997, Noise reduction and detection of weak,
  coherent signals through phase-weighted stacks: Geophysical Journal International, 
  {\bf 130}, 495--505.

\bibitem[Sinha et~al., 2005]{Sinha_S_2005_j-geophysics_spe_dsdcwt}
Sinha, S., P.~S. Routh, P.~D. Anno, and J.~P. Castagna,  2005, Spectral
  decomposition of seismic data with continuous-wavelet transform: Geophysics,
  {\bf 70}, P19--P25.

\bibitem[Spitz et~al., 2009]{Spitz_S_2009_p-eage-marine-w_sim_sswfmpefas}
Spitz, S., G. Hampson, and A. Pica,  2009, Simultaneous source separation using
  wave field modeling and {PEF} adaptive subtraction: EAGE Marine Seismic Workshop, M11.

\bibitem[Taner, 1980]{Taner_M_1980_j-geophys-prospect_lon_psfms}
Taner, M.~T.,  1980, Long period sea-floor multiples and their suppression:
  Geophysical Prospecting, {\bf 28}, 30--48.

\bibitem[Taner et~al., 1979]{Taner_M_1979_j-geophysics_com_sta}
Taner, M.~T., F. Koehler, and R.~E. Sheriff,  1979, Complex seismic trace
  analysis: Geophysics, {\bf 44}, 1041--1063.

\bibitem[Taner et~al., 1995]{Taner_M_1995_j-geophys-prospect_lon_pmspdxtd}
Taner, M.~T., R.~F. O'Doherty, and F. Koehler,  1995, Long period multiple
  suppression by predictive deconvolution in the $x-t$ domain: Geophysical
  Prospecting, {\bf 43}, 433--468.

\bibitem[Trad et~al., 2003]{Trad_D_2003_j-geophysics_lat_vsrt}
Trad, D., T. Ulrych, and M. Sacchi,  2003, Latest views of the sparse {Radon}
  transform: Geophysics, {\bf 68}, 386--399.
  
  \bibitem[van Groenestijn and Verschuur,
  2009]{VanGroenestijn_2009_j-geophysics_estimation_pnorsimda}
van Groenestijn, G. J.~A., and D.~J. Verschuur,  2009, Estimation of primaries
  and near-offset reconstruction by sparse inversion: Marine data applications:
  Geophysics, {\bf 74}, R119--R128.
  
  \bibitem[Ventosa et~al., 2011]{Ventosa_S_2011_p-eage_complex_wamsuf}
Ventosa, S.  H. Rabeson, P. Ricarte, and L. Duval, 2011, Complex wavelet adaptive multiple subtraction with unary filters: 73th Annual Meeting, EAGE, Expanded Abstracts.  
  

  \bibitem[Ventosa et~al., 2012]{Ventosa_S_2012_j-geophysics_window_lsosrlsst}
Ventosa, S., C. Simon, and M. Schimmel,  2012, Window length selection for an
  optimum slowness resolution of the local slant stack transform: Geophysics, {\bf 77}, V31--V40.

\bibitem[Verschuur and Berkhout, 1992]{Verschuur_D_1992_j-geophysics_ada_srme}
Verschuur, D., and A. Berkhout,  1992, Adaptive surface related multiple
  elimination: Geophysics, {\bf 57}, 1166--1177.

\bibitem[Verschuur, 2006]{Verschuur_D_2006_book_sei_mrtppf}
Verschuur, D.~J.,  2006, Seismic multiple removal techniques: past, present and
  future: EAGE Publications.

\bibitem[Verschuur and Berkhout,
  1997]{Verschuur_D_1997_j-geophysics_est_msiip2pae}
Verschuur, D.~J., and A.~J. Berkhout,  1997, Estimation of multiple scattering
  by iterative inversion, part {II}: Practical aspects and examples:
  Geophysics, {\bf 62}, 1596--1611.

\bibitem[Vetterli and Kova\v{c}evi\'{c}, 1995]{Vetterli_M_1995_book_wav_sc}
Vetterli, M. and J. Kova\v{c}evi\'{c},  1995, Wavelets and subband coding: Prentice-Hall.

\bibitem[Wang, 2003]{Wang_Y_2003_j-geophysics_mu_suemmf}
Wang, Y.,  2003, Multiple subtraction using an expanded multichannel matching
  filter: Geophysics, {\bf 68}, 346--354.

\bibitem[Weglein et~al., 1997]{Weglein_A_1997_j-geophysics_inv_ssmamsrd}
Weglein, A.~B., F.~A. Gasparotto, P.~M. Carvalho, and R.~H. Stolt,  1997, An
  inverse-scattering series method for attenuating multiples in seismic
  reflection data: Geophysics, {\bf 62}, 1975--1989.

\bibitem[Weisser et~al., 2006]{Weisser_T_2006_j-fb_wav_emmai3dsrme}
Weisser, T., A.~L. Pica, P. Herrmann, and R. Taylor,  2006, Wave equation
  multiple modelling: acquisition independent {3D SRME}: First Break, {\bf 24},
  75--79.

\bibitem[Wu and Wang, 2011]{Wu_M_2011_j-tle_cas_sfkd2dosd}
Wu, M., and S. Wang,  2011, A case study of $f-k$ demultiple on {2D} offshore
  seismic data: The Leading Edge, {\bf 30}, 446--450.

\bibitem[Yilmaz, 2001]{Yilmaz_O_2001_book_sei_dapiisd}
Yilmaz, {\"{O}}.,  2001, Seismic data analysis: processing, inversion, and
  interpretation of seismic data: SEG.

\end{thebibliography}

\end{document}